\documentclass[epjST]{svjour}
\usepackage{hyperref}
\usepackage{url}
\usepackage[square,numbers]{natbib}
\usepackage{graphics}
\newcommand{\comment}[1]{}
\def\greaterthansquiggle{\raise.3ex\hbox{$>$\kern-.75em\lower1ex\hbox{$\sim$}}}
\def\lessthansquiggle{\raise.3ex\hbox{$<$\kern-.75em\lower1ex\hbox{$\sim$}}}
\def\gl{\raise.3ex\hbox{$<$\kern-.68em\lower1ex\hbox{$>$}}}
\newcommand{\gts}{\greaterthansquiggle}

\newcommand{\be}{\begin{equation}}
\newcommand{\ee}{\end{equation}}
\newcommand{\met}{\mbox{$\not \hspace{-0.15cm} E\!\!_{T} \hspace{0.1cm}$}}

\hypersetup{pdfstartview=FitV,colorlinks=true,linkcolor=blue,citecolor=red,filecolor=black,urlcolor=blue}

\usepackage{amsmath,amssymb}

\newcommand{\newc}{\newcommand}
\newc{\wt}{\widetilde}
\def \met{\rm E{\!\!\!/}_T}

\def \ch2p {{\wt\chi_2^+}}

\def \ch2m {{\wt\chi_2^-}}

\def \chipm{{\wt\chi_i}^{\pm}}

\def \chonepm{{\wt\chi_1}^{\pm}}

\def \mchonepm{m_{\chonepm}}

\def \chtwopm{{\wt\chi_2}^{\pm}}

\def \lspi{\wt\chi_i^0}

\def \lspj{\wt\chi_j^0}

\def \lspone{\wt\chi_1^0}
\def \mlspone{m_{\lspone}}
\def \lsptwo{\wt\chi_2^0}

\def \lspthree{\wt\chi_3^0}

\def \lspfour{\wt\chi_4^0}

\def\omegapl{\Omega_{PL}}

\begin{document}
\title{Status of low mass LSP in SUSY.}
\author{Rahool Kumar Barman\inst{1}\fnmsep\thanks{\email{psrkb2284@iacs.res.in }}
Genevieve Belanger\inst{2}\fnmsep\thanks{\email{belanger@lapth.cnrs.fr}} \and Rohini M. Godbole\inst{3}\fnmsep\thanks{\email{rohini@iisc.ac.in}}
}
\institute{Indian Association for the cultivation of Science, 
Jadavpur, Kolkata 700032, India. \and LAPTh, Universit\'e Savoie Mont Blanc, CNRS, B.P. 110, F-74941 Annecy Cedex, France. \and Centre for High Energy Physics, Indian Institute of Science, Bangalore, 560012, India.}
\abstract{In this article we review the case for a light ($< m_{h_{125}}/2$) neutralino and sneutrino  being a viable Dark Matter (DM) candidate in Supersymmetry(SUSY).  To that end we recapitulate, very briefly, three issues related to the DM which impact the discussions : calculation of DM relic density,  detection of the DM in Direct and Indirect experiments and creation /detection at the Colliders. In case of SUSY,  the results from Higgs and SUSY searches at the colliders also have implications for the DM mass and couplings. In view of  the constraints coming from all these sources, the possibility of a light neutralino  is all but ruled out for the constrained MSSM : cMSSM. The pMSSM, where the gaugino masses are not related at high scale,  is also  quite constrained and under tension in case of thermal DM and will be put to very stern test in the near future in Direct Detection (DD) experiments as well as by the LHC analyses. However in the pMSSM with modified cosmology
and hence non thermal DM  or in the NMSSM, a light neutralino is much more easily accommodated. A light RH sneutrino  is  also still a viable  DM candidate although it requires  extending the MSSM with additional singlet neutrino superfields.  All of these possibilities can be indeed tested jointly in the upcoming  SUSY-electroweakino  and Higgs searches at the HL/HE luminosity LHC, the upcoming experiments for the Direct Detection (DD) and indirect detection for the DM as well as the high precision electron-positron colliders under planning.
} 
\maketitle
\section{Introduction}
\label{intro}
Particle physics finds itself currently  at a very curious juncture. It has been now almost a decade since the Higgs discovery~\cite{Zyla:2020zbs}.  Measurements of the various  properties of the Higgs such as its mass and couplings with all the SM particles are now available. The measurements are consistent with the observed state being SM-like.  The observation of a SM-like Higgs with a mass close to  the electroweak (EW) scale was expected to be accompanied by evidence for the TeV scale physics which would explain stabilisation of  the Higgs mass around the EW scale.  Absence of any evidence of such physics beyond the SM (BSM)  is what one can call the LHC paradox.  This situation is quite unusual in the field of particle physics where the development in the past six to seven decades had been particularly propelled by close cooperation/competition between the theory and experimental community where each set new goal posts for the other. 

Weak scale Supersymmetry (SUSY)~\cite{Drees:2004jm,Baer:2006rs} has been one of the  most attractive solution to the problem of the instability of the Higgs mass to radiative corrections  and hence had been a template of BSM physics to be searched  at the LHC. Absence of any evidence for any of the BSM physics at the TeV scale in the current LHC  results~\cite{Zyla:2020zbs}, 
 points toward generically  heavy and/or compressed supersymmetric spectra in the strong sector. The  limits are generically lower in the electroweakino (EWeakino) sector due to the lower EW production cross-sections as well as compressed spectra~\cite{Chan:1997bi,PhysRevLett.109.161802}. Experiments at the LHC have begun to probe such scenarios\comment{~\cite{PhysRevD.97.052010,Aad:2019qnd,Sirunyan:2018iwl}}.   The high luminosity LHC run, with an expected luminosity of $3000~{\rm fb^{-1}}$, will explore the EWeakino sector in detail.  In view of this and also the ever increasing lower limits on the strongly interacting sparticles, it is therefore a matter of great interest to investigate how light the EWeakino sector of the Supersymmetric theories can really be, remaining consistent with all the current available information and further what would be the prospects of current/future facilities to search for them.

Further the EW sector of supersymmetric theories involves 'naturally' a ready made candidate  for the Dark Matter (DM),  which is known to constitute about $26.8\%$ of the energy budget of the Universe, as extracted from the precise measurements of the Cosmic Microwave Background (CMB)~\cite{Aghanim:2018eyx}. Measurements by the PLANCK collaboration has put the DM relic density  to be 
\be
\omegapl h^2 = 0.120 \pm 0.001,
\label{eq:omegapl}
\ee
where  $h$ denotes the Hubble parameter in units of $100~km~s^{-1}~Mpc^{-1}$.  
Thus the upper limit on the $\Omega h^2$ at $95 \%$ C.L., $\Omega_{obs}^{max} h^{2}$ is $0.122$. The DM is one of the unsolved puzzles in the development of an understanding of the Universe and is naturally  at the forefront of theoretical and experimental research~\cite{Bertone:2010zza,Bertone:2016nfn}. The DM is indeed a problem looking for solution. While there exists no particle in the SM which can be an obvious candidate for the  DM, SUSY has a ready made 
DM candidate  in the form of the lightest supersymmetric particle, the LSP. The lightest neutralino or the sneutrino both have EW interactions (decided by their $SU(2)_L \times U(1)_Y$ quantum numbers) of appropriate strength to provide the correct order of magnitude of the relic density for masses of $\sim {\cal O} (100)~{\rm GeV}$, such that they are non relativistic at the time of 
freeze-out in the early universe~\cite{Kolb:1990vq}. Thus they are ideal WIMP (Weakly Interacting Massive Particle) DM candidates. Early comprehensive  discussions of SUSY DM can be found, for example, in \cite{Jungman:1995df,Griest:2000kj,Drees:2004jm}\footnote{For a recent mini review  of thermal and non thermal dark matter see~\cite{Dev:2013yza}. For a recent, detailed discussion of cosmological constraints on light, scalar dark matter see reference~\cite{Boehm:2020wbt}.}.

 A constraining and hence interesting  aspect of the SUSY DM solution is as follows.  The relic density predictions are obviously correlated with the  DM detection (direct/indirect) rates  and  DM searches at the  colliders as they involve the same couplings and masses, but interestingly enough they are also correlated with the  results of the searches for sparticles and Higgs bosons at the various colliders, studies of Higgs properties and precision measurements in flavour physics. This happens because the same parameters of the SUSY model which determine the mass of the DM particles and its couplings  with SM particles, also impact the spectrum and couplings of the Higgses for example.  As we will discuss in the next section in the simplest version of SUSY theories, the so called constrained Minimal Supersymmetric Standard Model (cMSSM), consistency with the  non-observation of SUSY particles at the LHC and absence of a signal in the direct/indirect  detection so far, coupled with the demand that the LSP DM has a relic density which can account for the DM in the Universe, all push the LSP mass to high values. A high mass LSP is in tension with the 'naturalness' of the SUSY solution. In addition, a low mass LSP can still be consistent with all the experimental constraints in other versions of MSSM such as the one with non universal Higgs masses (NUHM) or the phenomenological MSSM (pMSSM) where  the assumptions made to reduce the number of  parameters in the cMSSM are relaxed.
 
On the observational side, over more than a decade, there have been different claims of experimental evidence for 'light' DM in the mass range  $\sim {\cal O} (10)$ GeV~\cite{Gresham:2013mua}\footnote{We will summarise the current experimental situation in  some detail in the appropriate section.}. Further, the reach of major DD experiments deteriorates  in lower mass region and  many new ones are being constructed to probe even the sub-GeV region~\cite{Abdelhameed:2019hmk,Abdelhameed:2019szb}. All this indicates that the region of light LSP's is not without theoretical and observational interest. 

All the above makes it clear why investigating the status of a light LSP\footnote{We consider the region  $< m_{h125}/2$ for reasons to be explained later} which can be (at least partially) responsible for the relic DM in the Universe, is  very important. Within the context of SUSY, this is the one window that needs to be thoroughly explored, to be able to draw firm conclusions about the WIMP paradigm, which is under great scrutiny~\cite{Dev:2013yza,Bertone:2018krk}.

In this article we will mainly discuss the case of a low mass neutralino as the SUSY DM, and we will also mention briefly the status of sneutrino DM. In Section~\ref{sec:1} we will summarise the issues about the calculation of relic density of DM, its direct and indirect detection as well as some aspects of the collider phenomenology of the DM. After this we will discuss generic aspects of
the phenomenology of a light SUSY DM. We will show that a light neutralino LSP DM,  is all but ruled out in the most constrained SUSY models and is under tension in some of the variants. There are three ways to relax some of the tight constraints on the light LSP : 1) modifying the high scale relation among the gaugino masses or the Higgs mass parameters, 2) adding extra  EW particles (and hence EW sparticles) in the spectrum while keeping the same gauge group, e.g adding a singlet Higgs in the NMSSM and/or a right handed neutrino and 3) extending the gauge group. All three can lead to a change in the composition of the LSP and the latter two can change the number of interaction eigenstates of neutralinos and hence the nature of the LSP. Here we will discuss only the first two possibilities.

\section{WIMP and SUSY}
\label{sec:1}
\subsection{Relic density for a WIMP}
\label{subsec:2.1}
Let us begin here briefly by summarising the relevant details of the relic density calculation. In the standard model of cosmology, we assume that the early universe can be described by a radiation dominated medium that is in thermal equilibrium. Over time, the number density, $n$, of particles of a certain species within this medium depends on three rates: the expansion rate of the universe, the rate at which new particles are created, and the rate at
which they are annihilated. The time evolution of the number density can be described, under a few simplifying assumptions, by a Boltzmann evolution equation~\cite{Kolb:1990vq}:
\be
{\frac{dn}{dt}} = - 3Hn - \langle \sigma_{ann} v \rangle (n^2 - n_{eq}^2).
\label{eq:boltzman}
\ee
where $H$ is the Hubble parameter governing the expansion rate of the Universe,
$\langle \sigma_{ann} v \rangle$ is the velocity weighted annihilation cross-section  and $n_{eq}$ stands for the equilibrium number density. 
If the second term on the right-hand side dominates, $n$ will approach
the equilibrium number density $n_{eq}$. When $n$ reaches $n_{eq}$, the creation (proportional to $n_{eq}^2$) and annihilation (proportional to $n^2$) processes are in equilibrium, and the number density only changes over time because of the expansion of the universe. Therefore, the
Boltzmann equation will always drive towards an equilibrium.
During the inflationary era of the universe, the hot, dense medium of the early universe cools quickly and the number density of particles rapidly drops. Particle annihilations still occur during this stage, but the creation of heavier particle states becomes improbable because the thermal velocities of particles are too low. The number density therefore depletes. At some point, the density of the particles gets too low such that the annihilation mechanism also stops working: the universe is simply too dilute for particles to find each other. The creation and annihilation processes freeze out, and the number density approaches a constant value: the relic density $\Omega_{DM} h^2$. The relic density for a particular species would be decided by the thermal average $\langle \sigma_{ann} v \rangle$ and of course the mass of the DM candidate.  

In fact for a 'cold'\footnote{By Cold Dark Matter (CDM) one means that the DM particle is moving with non-relativistic speeds at the time of decoupling and freeze-out. If the DM particle is moving with relativistic speeds at the time of the freeze-out, it is called a Hot Dark Matter (HDM) candidate.} DM particle (call it $\widetilde \chi$), an approximate estimate of the relic density can be obtained  for standard thermal history of the Universe, which is independent of its mass, barring logarithmic corrections. The expression, neglecting these corrections, can be written as~\cite{Jungman:1995df}:

\be
\Omega_{\widetilde\chi} h^{2} = {\frac{m_{\widetilde\chi} n_{\widetilde\chi}}{\rho_c}} \simeq {\frac{3 \times 10^{-27} {\mathrm cm}^2 s^{-1}}{\langle \sigma_{ann} v\rangle}},
\label{eq:relic}
\ee
where $\rho_c$ is the critical density. 
The velocity dependence of the annihilation cross-section has been neglected in arriving at this. We see that if  the velocity averaged annihilation cross-section is  $\sim 10^{-26} {\mathrm cm}^2 s^{-1}$, one can have relic density with the right order of magnitude. This is 'naturally' achieved for a  weakly interacting massive particle with mass around $100$ GeV.  It is in this sense that the supersymmetric theories can present a ready made DM candidate which could have the correct relic density. Note, however,  that  this is also achieved  for a range of weakly interacting particle masses with the coupling strengths, $g_{\widetilde \chi}$,  varying appropriately such that  the ratio ${\displaystyle g_{\widetilde\chi}^2/m_{\widetilde\chi}}$ is kept fixed, because assuming $m_{\widetilde\chi}$ to be the only mass scale, on dimensional grounds we have ${\displaystyle \sigma_{ann} \sim g^4_{\widetilde \chi}/m^2_{\widetilde\chi}}$ . We will see effects of this in our discussions of SUSY DM as well.

Of course, most generally  $\langle \sigma v \rangle$ is  not velocity independent. On solving the Boltzman equation one finds that the freeze-out occurs roughly at a temperature $T_f$ such that $m_{\widetilde\chi} \simeq 20 -30 ~T_f$ if one wishes to have  the relic density $\Omega^{obs}_{\widetilde \chi} h^2 = 0.120 \pm 0.001$.  This means that $\widetilde \chi$ has small velocities at the freeze-out. Hence while calculating the relic density it is sufficient to expand the annihilation cross-section in powers of $v$ as
\be
\sigma v = a + b v^2 + \cdots .
\label{eq:vexpa}
\ee
For $s$-channel annihilation the cross-section will be constant and hence the first term can be enough. However, when $\widetilde \chi$ are Majorana particles, as they are in SUSY, the s-channel annihilation into light fermions, for example,  will be helicity suppressed and  the $b$ term HAS  to be taken into account as well. In general, it is sufficient to keep only the first two terms in the expansion and approximate analytical expressions can be obtained in terms of $a,b$~\cite{Jungman:1995df}. Of course, the state of the art relic density calculations are done by solving the Boltzman equation numerically and considering all the annihilation channels~\cite{Belanger:2001fz,Belanger:2004yn,Belanger:2014vza,Belanger:2018mqt}. Note that if the annihilation proceeds through a resonance then these rules of thumb are not enough to understand the results one gets numerically. 

If for a particular DM species, say $\widetilde \chi$, the computed relic density
 $\Omega_{\widetilde \chi} h^2$ is  less than the observed one $\omegapl h^2$  then that species is said to be under-abundant, as it does not account for all the DM that is observed. Likewise, if one has $\Omega_{\widetilde \chi} >  \omegapl $ the corresponding  DM species is said to be over-abundant. Recall here also that while the velocities of the DM particles at the freeze-out are $\sim 0.1-0.3 c$, those in the galactic halos are much smaller$\sim 200~{\mathrm km}~ {\mathrm s}^{-1} \sim 10^{-3} c$. This will have to be kept in mind when one tries to understand in a collective manner the implications of  parameters of a model for the  relic density  as well as  those for the direct/indirect detection experiments in a given model.  It should also make it clear why the issue of velocity dependence of $\langle \sigma_{ann} v \rangle$ is a crucial one.
 
It is also possible, as in Supersymmetric theories, that there exist other BSM particles in the spectrum. If there exists the next to lightest supersymmetric particle (NLSP) (say $\widetilde \chi_1$) which is close in mass to the LSP (which we have called $\widetilde \chi$), then in addition to the self annihilation via $\widetilde \chi + \widetilde \chi \rightarrow  SM + SM$, $\widetilde\chi$ can be depleted also by the co-annihilation process $\widetilde \chi + \widetilde \chi_{1} \rightarrow SM + SM$. If the masses of $\widetilde \chi$ and $\widetilde \chi_{1}$ are close then the number density of $\widetilde \chi_{1}$ is appreciable at the time $\widetilde \chi$ freezes out. Then this depletion needs to be added in the Boltzman equation(Eq.~\ref{eq:boltzman}). 

Further sometimes the LSP interacts so weakly with the SM particles that it cannot reach thermal  equilibrium. In this case DM can be created either from the decay of some heavy particle that decays into the LSP while in equilibrium or from the annihilation of pair of SM particles. This is called the freeze-in mechanism~\cite{Hall:2009bx,McDonald:2001vt}.

Another possibility is that the NLSP decays on a long time scale, either due to compressed spectra or small couplings, into a final state containing the LSP. When this decay occurs after the freeze-out of the NLSP,  the relic density of the LSP is simply related to that of the NLSP

There exists of course the possibility that the early universe cosmology is non standard. In the above discussion, one has assumed that the Universe must have evolved adiabatically after the $\widetilde \chi$  decoupled. 
If there was a period of entropy production, e.g. due to
the out--of--equilibrium decay of another massive particle, only a
small fraction of today's CMB photons would originate from the SM
plasma at $T_f$ which is the freeze-out temperature. Only this
fraction should be used in computing  $n_{\widetilde \chi}$  and then the prediction of $\Omega_{\widetilde\chi} h^2$ will be diluted accordingly.  One example of such a late decaying particle, is a SUSY modulus scalar. For every value 
of DM mass, several combinations of reheating temperature and heavy scalar mass can lead to  a relic density compatible with the observed value~\cite{Gelmini:2006pw,Gelmini:2006pq,Hooper:2013nia}. One phenomenological approach to analyse the over-abundnant scenarios would therefore be to  simply assume that it is possible to find a  mechanism that brings  the DM relic density in agreement with 
observations. Thus, in practice one can analyse all the parameter space 
points of the model, for which  the relic density value computed assuming thermal freeze-out with a standard cosmological model is above $\Omega_{PL} h^{2}$ and ask the question how these parameter space points may be explored by the direct/indirect detection experiments or the collider ones. 

In the next two sub sections we first discuss the  (model independent) aspects of all the three types of DM detection experiments and then discuss their interpretation as well as implication  in the context of SUSY model parameters in the last subsection.

\subsection{Direct and Indirect Detection of DM}
Let us briefly summarise the (model independent) information about Direct and Indirect detection of the DM.  Direct detection experiments aim to detect small perturbations of atoms within the detectors, which are caused by WIMPs of astrophysical origin that pass the detector. Since the velocity of these particles is generally non-relativistic, the WIMP scattering occurs elastically and at most causes excitation or ionization of detector material. As we do not know the mass, the type of interaction, or the interaction strength of the WIMPs, it is important to have several detectors with different detector materials. Furthermore, to reduce the background as much as possible, direct detection experiments are often placed in deep-underground laboratories. 
The goal of these highly sensitive experiments is to measure the amount of energy deposited when a WIMP DM scatters off the target nuclei inside the detector in a background-free environment. The scattering rate between the WIMP and target nuclei ($dN/dE$) can be obtained~\cite{Goodman:1984dc,Queiroz:2016awc,Duda:2006uk}.
This of course depends on the WIMP-nuclei differential cross section which, in turn depends, among other things, on the distribution of the WIMP DM particles in the relative velocity $v$ between the DM and the earth. The interaction cross-section has two parts: spin-independent $\sigma_{SI}$ and the spin-dependent $\sigma_{SD}$. Note that the spin-independent interactions are coherent as they couple to the entire nucleus whereas the spin dependent interactions are not, because the spin of the nucleus does not increase with its mass. Hence, for heavier nuclei the spin-independent interactions dominate. The calculation  requires knowledge of the spin content of the nuclei as well as degree of coherence between different nucleons. The latter is encoded in the form factors calculated in \cite{Duda:2006uk}.
\begin{figure}
\begin{center}
\resizebox{0.60\columnwidth}{!}{
\includegraphics{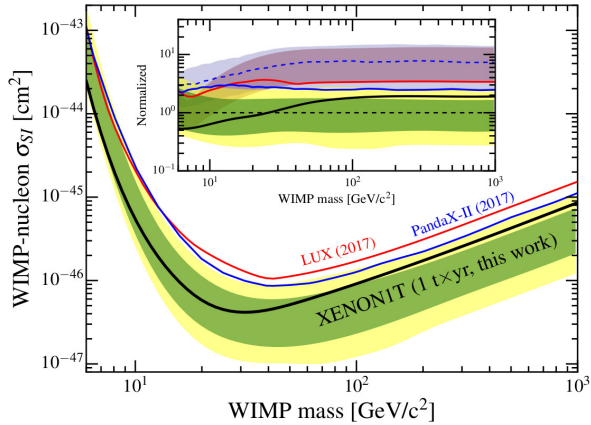} }
\caption{Upper limits on the spin-independent WIMP-nucleon cross section ($\sigma_{SI}$), derived by Xenon-1T, at $90\%$ C.L.~(from \cite{Aprile:2018dbl}).}
\label{Fig:DM_current}
\end{center}
\end{figure}
The non-observation of WIMPs at direct detection experiments places limits on the interaction strength between a WIMP and the proton/neutron as a function of the WIMP mass. Fig.~\ref{Fig:DM_current}~(taken from \cite{Aprile:2018dbl}) summarises the current upper limits on the spin-independent WIMP-nucleon cross section from the Xenon-1T experiment.  Note also that even though there exist considerable uncertainties  in the theoretical  predictions of the expected cross-sections, coming from astrophysical inputs as well as limitations in the knowledge of the parton  content of the nucleus, they are under reasonable control~\cite{Goodman:1984dc,Queiroz:2016awc}. The DD detectors have a threshold energy, below which they cannot measure the recoil induced by the DM particle. This means that the DD detectors have low sensitivity for low-mass DM particles. As the mass of the DM particle becomes higher, the number density and therefore the flux decreases, so we expect a lower rate. This explains the observed shape of the limits on the WIMP-nucleus cross-section seen in Fig.~\ref{Fig:DM_current}. In fact, below ${\cal O} (10)$ GeV (depending on the detector material) there is low sensitivity. Above ${\cal O} (10)$  GeV, the sensitivity increases till about ${\cal O} (50 )$ GeV and then reduces again due to the limited statistics for high-mass DM particles.

A short comment on the evidence for light WIMP (in the mass range from 7 to 30 GeV) that was mentioned in the introduction, is in order here. The results from LUX~\cite{Akerib:2016vxi} and Xenon-1T~\cite{Aprile:2018dbl} cover, with much higher sensitivity,  the same region where  signals were reported by CoGenT~\cite{Aalseth:2010vx}, CRESST~\cite{Angloher:2011uu}, CDMS-Si~\cite{Ahmed:2009zw,Agnese:2013rvf} and DAMA/LIBRA~\cite{Bernabei:2010mq,Bernabei:2014jnz}. The absence of signal by these two experiments in this region implies that the earlier experimental results are either fluctuations or not related to DM. Results by an improved version of CDMS, the SuperCDMS, does not see any signal in the region favoured by earlier sightings either~\cite{Agnese:2014aze}. It should be noted here that the region of small DM masses (smaller than $6$ GeV or so ) is actually quite unconstrained even after the very remarkable  result from Xenon-1T, shown in Fig.~\ref{Fig:DM_current}, has become available. It is an interesting question to ask that if a light $\lspone$ is detected and  if one manages to construct a model such that  it is allowed from relic density considerations,
whether other cosmological considerations will allow it. This was answered in the affirmative~\cite{Boehm:2002yz}. Using the bound on effective number of $\nu$ species ($N_{eff}$), one can in fact show that, on cosmological grounds even a Cold Dark Matter (CDM) of mass $\sim {O}$ (MeV) is allowed~\cite{Boehm:2013jpa}.

Indirect detection experiments look for annihilation products that originate from astrophysical WIMP--WIMP scattering. These products include (anti-)protons, photons, (anti-)electrons, and neutrinos. The rate at which these particles are created depends on the density of dark matter squared. Furthermore, the velocities of the WIMPs need to be high enough, such that the WIMPs can scatter inelastically. These facts combined mean that these annihilations occur in dense areas, such as the dwarf spheroidal (dSph) galaxies which are rich in DM or the center of the sun or the galaxy. The non-observation of the annihilation products results in limits on the present-day velocity-weighted annihilation cross section. These limits depend on the WIMP mass and the annihilation scenario. However, WIMP--WIMP scattering is not the only process that could create (anti-)protons, photons, (anti-)electrons or neutrinos. It follows that a good knowledge of the environments where the annihilations could occur
is needed, and this knowledge is not always available. An example of the foregoing is the notorious Galactic Center excess of the spectrum of high energetic photons (gamma rays), as observed by the Fermi-LAT satellite~\cite{TheFermi-LAT:2017vmf}. A possible explanation for such excesses includes annihilations of light ($< 60-70$ GeV) DM particles, see for example, ~\cite{Goodenough:2009gk,Hooper:2010mq,Hooper:2011ti,Daylan:2014rsa}.
But given the many uncertainties, these explanations are heavily debated. 

\subsection{DM at colliders}
The dark matter particles are assumed to interact weakly with the observed matter, are stable at the time scale of the universe, and, are assumed to be charge neutral. Correspondingly, the DM particles manifest themselves as missing energy in the colliders. Hence the collider-based experiments, such as ATLAS and CMS, look for missing transverse energy ($\met$) in their detectors, which may be a sign of an undetected particle that is produced during the inelastic scattering process of the two colliding protons.  Dark matter searches at the colliders are therefore mostly based on identifying the visible counterparts produced in association with the DM candidate $viz$ mono-jet$~+~\met$~\cite{Sirunyan:2017hci,Aaboud:2017phn}
, mono-$Z/W^{\pm}/H~+~\met$~\cite{Sirunyan:2018gdw,Aaboud:2017yqz}. 
DM particles can also be present in decay chain of sparticles, should they be produced in the $pp$ collisions. These then can give rise to events containing $\met$ along with SM particles. This happens to be an important channel for SUSY searches as well. None of these searches have reported a clear signature over the SM expectation at the LHC so far.  
 
Another category of collider probes for the case of light DM ($m_{\widetilde\chi} \leq m_{h_{125}}/2$) are the direct searches of invisibly decaying Higgs bosons. These searches typically consider a DM particle with mass below $\sim 62.5~{\rm GeV}$, such that it is kinematically feasible for the Higgs to decay into the DM candidates. A number of studies have analysed the prospect of probing the  invisible decays of the Higgs at the LHC,
e.g 
\cite{Gunion:1993jf,Choudhury:1993hv,Eboli:2000ze,Godbole:2003it}. 
In fact, both CMS and ATLAS have looked, in both Run-I and Run-II of LHC, for such invisibly decaying Higgs through its inclusive production in the $ggF$, $VBF$ and $Vh$ modes. 
The most recent measurement of $Br(h \to invisible)$, performed by ATLAS using the Run-II LHC data ~($\mathcal{L}=140~{\rm fb^{-1}}$) in the $VBF$ Higgs production mode, has set the upper limit at $13\%$ at $95\%$ CL~\cite{ATLAS:2020cjb}. Note here that the observed Higgs signal strengths also imply an direct, albeit model dependent, constraint on this branching ratio.

In the context of a specific model like SUSY the DM may also be searched at the collider in terms of production of just the EWeakinos which we will discuss in detail in the next sections.

An important aspect regarding the collider searches of DM is that any neutral particle which interacts weakly with the detector or which decays outside the detector would also result in a missing energy signature similar to that of a DM particle. Under such circumstances, it would not be feasible to resolve the DM contribution to $\met$. Therefore, the collider searches for DM of this variety do not provide the most efficient probes for DM detection, rather, they provide complementary probes in association with the direct and indirect DM detection modes.

\subsection{SUSY DM}
\label{subsec:susy_DM}
In the MSSM there are two neutral LSP candidates in the observable sector, the lightest neutralino $\lspone$ and a $\widetilde\nu$. While in most models the $\lspone$ is the LSP we will briefly entertain the possibility of the $\widetilde \nu$ DM. We will not discuss the case of the light gravitino as the DM in this article.

Due to the absence of positive results in the  search for DM at the colliders as well as in the high sensitivity DM experiments, for a light $\lspone$, the focus shifted away from the constrained SUSY model cMSSM with its  limited number of parameters, quite early on. The interest, since then, has been in SUSY models where the constraining assumptions are relaxed or the particle content is augmented, with a view to obtain a SUSY DM candidate which 1)can provide adequate relic DM either via freeze-out or freeze-in and 2)is consistent with the non-observation in the DD experiments. The tight connection between  the DD rates and the  size of the $\sigma_{ann} v$ makes this exercise more tricky. The focus of the activity is to see how one can explore these models at the LHC not just through the DM searches but also via associated phenomenology of the Higgs and also that of the sparticles other than the LSP. Of course, the latter has meaning only in the context of a particular model.

\subsubsection{The sneutrino $\widetilde \nu$ DM}
Even though $\lspone$ is the LSP in most SUSY models, let us briefly discuss the possibility of a $\widetilde\nu$ LSP. In this case the most important
contributions to the annihilation cross section come from (1)~$\widetilde{\nu} \widetilde{\bar\nu} \rightarrow f \bar{f}$ through the exchange of a $Z$ boson in the $s-$channel, where $f$ is an SM fermion, (2)~$\widetilde{\nu} \widetilde{\bar\nu} \rightarrow l \bar{l}$
through the exchange of a neutralino or chargino in the $t-$channel
and (3)~$\widetilde{\nu} \widetilde{\bar\nu} \rightarrow V \bar{V}$ ($V = W^\pm, Z$) for $m_{\widetilde \nu} \geq m_V$. With the exception of the $l \bar{l}$ final states other final states do not require exchange of any particles that are possibly heavy. The relevant couplings are electroweak gauge couplings, unsuppressed by any mixing angles. Thus the annihilation cross-sections are rather large and hence by Eq. \ref{eq:relic} the resulting relic is small unless the $\widetilde\nu$ is quite heavy. Moreover, direct dark matter searches exclude the possibility that $\widetilde \nu$'s form a major part of the dark halo of our galaxy~\cite{Jungman:1995df}. The reason is again the full strength couplings of $\widetilde \nu$ with the $Z$ boson, which leads to a large scattering cross-section, bigger by a factor $4$ compared to a Dirac neutrino. Nevertheless the sneutrino could be the LSP. In this case the only {\it model independent} limit on $m_{\tilde\nu_L}$ comes from the invisible width of the $Z$ and is 41 GeV as quoted in reference~\cite{Zyla:2020zbs} and obtained in ~\cite{Decamp:1991uy}. Since the $\wt\nu_L$ couples only to the $Z$, this limit could not be improved by the 'mono' photon analysis at LEP-II with its limited luminosity. Situation can be different at the future $e^+e^-$ colliders.\footnote{The limits of $84$ GeV~\cite{Heister:2002jca} and $94$ GeV~\cite{Abdallah:2003xe} on the mass of $\wt\nu_L$, obtained using an analysis of slepton production by the ALEPH and DELPHI collaborations respectively, are valid for the case of the LSP being $\lspone$ and more over are in the framework of the cMSSM.} Above mentioned direct detection constraints can be avoided if one considers additional DM candidates, as was revisited recently in reference~\cite{Carpenter:2020fnh} where it was also shown that  some parameter space may be made viable by considering inelastic DM in the mixed $\widetilde \nu$ scenario. 

 Within this scenario, after recasting the current analysis of the mono-W/Z from the LHC Run-II and coupling this with  the upper bound on the invisible width of the Higgs boson, reference~\cite{Carpenter:2020fnh}
obtains a lower limit of $55$~GeV on $m_{\widetilde \nu}$, assuming a compressed spectrum with $\Delta m = m_{\widetilde e} - m_{\widetilde \nu} = 5$~GeV. In the absence of a compressed spectra, the $\widetilde \nu$
cannot be light. Moreover the light sneutrino scenario is expected to be fully covered by the HL-LHC assuming that the leptons for the $\widetilde e$ decay will be 'invisible', ie. for the same compressed spectra. Another option, that we will not entertain here, is that the sneutrino is the NLSP and DM consists of a  gravitino or an axino.
It is not clear, however, whether any regions of the parameter space of different models described in Refs. 42 to 54 in ~\cite{Carpenter:2020fnh} still remain viable to  keep  $\widetilde\nu$ as the LSP since both the collider constraints as well as cosmological ones have changed substantially since  the original models were constructed. The issue needs to be revisited too!

Adding a right handed (RH) neutrino and sneutrino $\widetilde \nu_R$ provides a well motivated extension of minimal SUSY that allows to account for neutrino masses. This also enlarges the possibilities for sneutrino DM. Indeed the DM could now be the  ${\widetilde \nu}_R$ or a mixed state. In the latter case one will still  suffer from strong bounds from direct detection that enters through the small ${\widetilde\nu}_L$ component. Those can be best avoided if the $\widetilde\nu$ LSP  lies in the region where DD experiments loose sensitivity (less than 10 GeV)~\cite{Dumont:2012ee}. \comment{Considerations of the $N^{eff}$ limits mentioned before, were used to find a lower limit of $3.5$ MeV on $\widetilde\chi_1^0$ as CDM~\cite{Boehm:2013jpa}.}  Obtaining a  light mixed sneutrino, however, requires a very large $A$ term in the sneutrino mass matrix. This is hard to achieve in the pMSSM with parameters defined at the high scale. Thus in this framework the light sneutrino has been shown to be no longer viable after taking into account LHC constraints ~\cite{Arina:2015uea}.

The case of a pure ${\widetilde \nu}_R$ has been discussed within various models,  for example the
 cMSSM, the pMSSM  ~\cite{Banerjee:2016uyt,Banerjee:2018uut} or the
NMSSM~\cite{Cerdeno:2008ep,Cerdeno:2014cda,Cerdeno:2015ega} extended with ${\widetilde\nu}_R$.
 In the MSSM one adds three generations of right handed singlet $\widetilde \nu$ and the superpotential  can be written as
\be
{\cal W}_{\widetilde \nu mssm} = {\cal W}_{mssm} + y_\nu \hat H_u \cdot  \hat L \hat N.
\label{eq:snuMSSM}
\ee
where $N$ is a the singlet neutrino field and $y_\nu$ is the Yukawa coupling producing Dirac masses for the $\nu$'s.  The only interaction between the  ${\widetilde\nu}_R$ and the SM particles is controlled by the small $y_\nu$.  This smallness  has various implications. Firstly the very weak interaction allows to completely avoid the strong constraints from  DD cross-sections. Moreover  it means that the ${\widetilde\nu}_R$ might not achieve  thermal equilibrium in the early Universe, two mechanisms can then contribute to the  relic  formation, the freeze-in production of $\widetilde\nu_R$ through the decay of some SM or SUSY particle, or the production of $\widetilde\nu_R$ from the decay of the long-lived NLSP after it freezes out.
 As a result of the small $y_\nu$, after the RG evolutions the $\widetilde \nu_R$ emerges quite often as the LSP and $\widetilde \tau_1$ as the NLSP. The $\widetilde \nu_L$ also can be quite often light but does not contribute to the relic DM in  any appreciable manner for reasons explained in the first paragraph. 
In ~\cite{Asaka:2005cn}, it is shown that it is possible to get $\Omega_{\widetilde \nu_R} h^2$  for $m_{\widetilde \nu_R} \sim 30 -40$ GeV, in agreement with the measurement by PLANCK collaboration~\cite{Aghanim:2018eyx}, with all the strongly interacting sparticles in the TeV range  and the EW sparticles in the range $\sim 500-1000$ GeV, consistent with the current limits from the LHC. Of course, as was already mentioned in the introduction, one has to make sure that the life time of the long lived $\widetilde \tau_1$ does not create problems with Big Bang Nucleosynthesis (BBN), that its mass is above the limit imposed on the quasi-stable charged particle from LHC searches and that  the mass of the light Higgs as well as the rates observed in various channels are consistent with measurements~\cite{Aad:2015zhl,Aad_2016}. This just shows the tight rope one has to walk before a SUSY DM candidate is acceptable. These are the correlations that were talked about in the introduction.

A light  $\widetilde \nu_R$ can also behave as a WIMP if it couples to other new particles, in that case it could  have detectable cross-section in DD experiments,  such is the case for example is an  extension of the NMSSM with the superpotential
\be
{\cal W}_{\widetilde\nu nmssm} = {\cal W}_{nmssm} + \lambda_N \hat S \hat N \hat N + y_\nu \hat L \cdot \hat H_u \hat N,
\label{eq:wsnnmssm}
\ee
where 
\begin{equation}
{\cal W}_{nmssm}  = {\cal W}_{mssm} (\mu = 0) + \lambda \hat{S}\hat{H_{u}}\cdot\hat{H_{d}} + \frac{\kappa}{3}\hat{S}^{3}
\label{eq:wnmssm}
\end{equation}
and $W_{mssm}$ ($\mu= 0$) refers to the MSSM super potential without the $\mu$-term, while $\lambda$ and $\kappa$ are dimensionless parameters.
Thus, compared with the NMSSM the model contains the trilinear interaction between the singlet superfield S and  one more additional singlet Neutrino super field $N$. \comment{In addition to the parameters $\lambda,~\kappa,\lambda_N$ mentioned in Eqs. \ref{eq:wsnnmssm} and \ref{eq:wnmssm}, the model also has the corresponding soft trilinear terms $A_{\lambda}, A_{\kappa},A_{\lambda_N}$ and the soft mass term for $\widetilde \nu_R$ as parameters.}Here  again the Yukawa coupling  $y_\nu$ is a rather small number as it gives rise to the small Dirac neutrino mass term. But this interaction does not play any role, rather it is the trilinear  interaction between the singlet fields $S$ and $N$ which gives rise to  interactions of the $\widetilde \nu_R$ with the Higgs sector and hence with the SM particles. Thus,
it is possible to have a $\widetilde \nu_R$ as a satisfactory thermal DM candidate over a wide range of parameters and in particular  to have a light DM candidate~\cite{Cerdeno:2008ep}. Let us just mention in the end that the invisible decays of the $h$ play an important role in the light $\widetilde\nu_R$ DM phenomenology mentioned and  has been studied, eg. in \cite{Banerjee:2013fga}. 

\subsubsection{The neutralino $\lspone$ DM}
Next we move to the much more widely studied case of the LSP neutralino DM.
To discuss this let us just summarise the parameter choice that is used normally for phenomenological discussions. Assuming no CP violation other than the one in the SM,  in the framework of pMSSM one has in fact 19 parameters, defined at the EW scale.  These are: the gaugino masses $M_1, M_2, M_3$, the Higgs sector parameters $\mu, \tan \beta= v_{u}/v_{d}, m_A$, the masses of the first two generations of sfermions  $m_{\tilde e_R}, m_{\tilde L_1}, m_{\tilde Q_1}, m_{\tilde u_R},m_{\tilde d_R}$,
those of the third generation $m_{\tilde \tau_R}, m_{\tilde L_3}, m_{\tilde Q_3}, m_{\tilde b_R}, m_{\tilde t_R}$ and the trilinear couplings $A_t, A_\tau, A_b$. 
 For the case of cMSSM, one has only four free parameters :
the common scalar mass $m_0$, the common gaugino  mass $M_{1/2}$, a common trilinear term $A_0$, Higgs sector parameters $\tan \beta$ and sign of $\mu$. Values of $\mu, m_A$ and all the other above mentioned parameters at the EW scale are  then calculated in terms of these high scale parameters. In the non-universal gaugino models the equality of gaugino masses at high scale is broken and all the other remains the same as cMSSM.  

The neutralino and chargino  mass matrix can be written in terms of the pMSSM parameters as:

\begin{small}
\begin{equation}
\mathcal{M}^{n}=
  \begin{pmatrix}
    M_{1} & 0 &-M_Zc_\beta s_W&M_Z s_\beta s_W \\
    0&M_2&M_Z c_\beta c_W&-M_Z s_\beta c_W \\
    -M_Z c_\beta s_W&M_Z c_\beta c_W&0&-\mu \\
     M_Z s_\beta s_W&-M_Z s_\beta c_W&-\mu&0 \\
    \label{eq:9p24}
  \end{pmatrix},
\end{equation}
\end{small}
where $s_W\equiv \sin\theta_W, c_W\equiv \cos\theta_W, s_\beta \equiv
\sin\beta$, $c_\beta\equiv \cos\beta$ and 
\begin{small}
\begin{equation}
{\bf X}=
  \begin{pmatrix}
    M_2 & \sqrt{2} M_W \sin\beta  \\
    \sqrt{2} M_W \cos\beta & \mu \\
    \label{eq:9p7}
  \end{pmatrix}.
\end{equation}
\end{small}
The neutralino mass matrix ${\cal M}^n$ is written in the interaction eigenstate ($\tilde B,\tilde W_3,\tilde h_d^0,\tilde h_u^0$) basis, ie. the bino-wino-higgsino basis and the chargino mass matrix is written in the wino-higgsino basis. The mass eigenstates of these two matrices, denoted by $\widetilde \chi_i^0, \widetilde \chi_j^\pm$ with $i=1,4$ and $j=1,2$ are mixtures of  gauginos and higgsinos, the mass increasing with increasing index. Thus $\lspone$ is one choice for the LSP. Note from Eqs.~\ref{eq:9p24}, \ref{eq:9p7}, that the masses and the gaugino-higgsino  mixing of the EWeakinos are controlled by $M_1,M_2, \tan\beta$ and $\mu$. The effect of $\tan \beta$ is somewhat mild. Thus the dominant component of the lightest EWeakinos is decided by the relative values of  $M_1,M_2,\mu$ : higgsinos if $|\mu| \ll M_1,M_2$ and bino (wino) for $M_1 (M_2) \ll |\mu|$. If $\mu, M_1$ and $M_2$ are all comparable then $\widetilde \chi_i^0, \widetilde \chi_j^\pm, i=1,4$ and $j=1,2$, are mixed states. In the cMSSM the $\lspone$ is dominantly a bino. 

The couplings of the various EWeakino mass eigenstates with the SM particles, are controlled by the gaugino/higgsino content. The annihilation cross-sections are controlled by these as well as masses of the sparticles exchanged in the $t$-channel or the gauge bosons/higgses in the $s$-channel. Note that the masses of the various EWeakinos and sfermions and the Higgses are controlled by the parameters in the list of the $19$ parameters mentioned for the pMSSM, whereas in the other versions of SUSY models the values of these at the EW scale will be decided in terms of the few parameters given at the high scale. Thus the dominant annihilation channel for a given type of $\lspone$ as well as whether there will be co-annihilations, all will be decided by these parameters. 

Thus it is no surprise that, while the discussions after Eq.~\ref{eq:relic} showed that SUSY DM $\lspone$ can have interactions and mass of the right order of magnitude required to give rise to the observed relic density $\omegapl h^2$, in reality as we scan over the parameter space of the SUSY models, the predictions for the relic density can vary by many orders of magnitude, controlled mainly, though not  completely, by the LSP composition. The subject is complex and the literature on the subject truly vast. We refer the reader to \cite{Roszkowski:2017nbc} for a recent summary. However, the general strategy that has been chosen is to extract the essential features of the expected relic and the annihilation as well as interaction cross-sections, in terms of the gaugino/higgsino content of the LSP. In each case, only a few of the many parameters are relevant and one can discuss the issue comprehensively and completely in terms of only those. We can specialise then the discussion to the case of interest here, viz. the light LSP. 

Pure winos~(higgsinos) annihilate readily into gauge boson pairs and thus have large $\sigma_{ann}$ and hence are required to be very heavy $\sim 2.5~(1)$ TeV, if they are to explain the observed relic $\Omega_{PL} h^{2}$. Reference~\cite{Chakraborti:2017dpu}, for example, points out model parameter regions where the limits can be brought down upto a factor 2, but still not enough to render them 'light' enough for our consideration. Thus clearly these can not play very important role for the  discussions of the light $\lspone$ DM.  

On the other hand, let us  take the case of a pure bino. The small size of the $U(1)_Y$ gauge coupling generally makes the annihilation cross-section small. This means that the $\lspone$ will freeze out very early and hence the relic density will be too high. The strength of the cross-section also depends on  the masses of sparticles exchanged in the $t/u$ channel for the production of pairs of longitudinal gauge bosons and $f \bar f$ pairs etc. Co-annihilation with sfermions can also provide some additional annihilation cross-section and decrease the relic density of binos. Hence these scenarios can then have implications for (and are impacted by), both the LHC searches and  the DD experiments. It is also not surprising that the LEP limits and the early days of LHC searches which put limits on the masses of the charged EW sparticles like $\tilde l_{iR}, \tilde l_{iL}$  and $\tilde \chi_{1,2}^\pm$,
already constrained considerably the viability of a pure light bino as a DM relic capable of explaining the observed relic density. However, the situation can change substantially once the $\lspone$ is of mixed nature, and even a small higgsino-bino mixing can increase annihilation cross-section through scalars $h,H$  or the pseudoscalar $A$, in the $s$-channel. The annihilation into a $f \bar f$ pair by $A$ resonance takes place via a $s$-wave, due to the Majorana nature of the $\lspone$ while the annihilation by the scalars $h,H$ happens via a $p$-wave.

From the above discussions, it is clear that in MSSM (and its variant) 
the possibility of having a light DM can be realised with a bino dominated $\lspone$. Since such a $\lspone$ was naturally expected in the low scale 'natural' SUSY, even before the various claims for light DM detection came on the scene, there was, in fact, a lot of interest in a light neutralino DM. Cosmological considerations  implied rather small lower bounds on the DM mass of ${\cal O}~(few)~{\rm GeV}$~\cite{Boehm:2002yz}. Discussions of~\cite{Boehm:2013jpa}
specialised to SUSY, also indicated a limit of $3.5$ MeV on 
$m_{\lspone}$.\footnote{In the MSSM extended to include a $\widetilde \nu_R$ which mixes with $\widetilde \nu_L$, considerations of the $N^{eff}$ limits mentioned before, were used to find a lower limit of $3.5$ MeV on $\lspone$ as CDM~\cite{Boehm:2013jpa}. Here the mixed $\widetilde \nu_R$ acts as an mediator for the annihilation of $\lspone$  to $\nu \bar \nu$. The Majorana nature of the $\lspone$ make the annihilation to be a $p$ wave process and hence there is no danger of distortion of the CMB radiation due to energy injection from the annihilation process.} In view of these lower bounds and also a light dominant bino being an excellent thermal DM candidate, focus shifted to a general MSSM framework which consisted of various versions of the pMSSM 
once the LEP bound on the mass of a $\widetilde\chi^\pm$ implied a lower bound on $\lspone$ of $46$ GeV in the cMSSM. The different versions of the pMSSM differed in the choice of free parameters, mainly relaxing the hypothesis of unification of gaugino masses in different ways. With the right thermal relic for a (dominantly) bino being facilitated by light slepton masses, this  even had the potential of explaining the DM as well as the $(g-2)_\mu$ at one stroke and this was a favoured SUSY scenario considered in the early days by 
many~\cite{Belanger:2000tg,Belanger:2001am,Hooper:2002nq,Belanger:2003wb}  to quote a few. In the analysis of \cite{Belanger:2003wb}  which made use of a existence of the pseudoscalar $A$ and/or a light slepton, a lower bound was obtained on the neutralino mass between $4$-$30$ GeV, depending on values of $m_A$ and $\tan \beta$.  This was consistent with the limit of about $18$ GeV of reference~\cite{Hooper:2002nq} which was obtained in a somewhat different scenario. It was shown that even an almost massless $\lspone$ is allowed by the then available collider and cosmological data, as well as precision measurements of meson decays~\cite{Dreiner:2009ic}, if the relic was due to a HDM LSP, a hypothesis that is not possible to sustain in view of the precision CMB measurements. The driving force on the constraints on $\lspone$ masses in this period were the collider experiments and considerations of correct relic density setting goal posts for DD experiments. It should be noted here that the contribution of a pseudoscalar  to the DM-nucleon scattering cross section is suppressed by the small momentum transfer. Hence  DD experiments are much more sensitive to a light scalar. Consequently they constrain a light scalar much more strongly than the pseudoscalar, for the same coupling and mass.

With  various claims of observation of light DM, both in direct and indirect detection experiments, the interest in light DM phenomenology really escalated with a large number of investigations looking at the viability of light bino-dominated $\lspone$ DM, with light sleptons particularly, $\widetilde \tau_R$,  with or without the resonant contribution of the $h/H/A/Z$~\cite{Kuflik:2010ah,Belikov:2010yi,Vasquez:2010ru,Calibbi:2011ug,Choi:2011vv,AlbornozVasquez:2011yq}.  
An analysis in the context of MSSM in~\cite{Vasquez:2010ru} put a lower limit of $28$ GeV in the MSSM, which satisfied all the collider and flavour physics constraints available then. The importance of the constraints from $b \rightarrow s \mu^+ \mu^-$ on the $m_A, \tan\beta$ values in doing this analysis and for the limits on the mass of the $\lspone$ was pointed out in~\cite{Feldman:2010ke}.  The possibility of LHC signals that one may search for in case of the light $\wt\tau_R$ which helps  a light bino like $\lspone$ good thermal DM, had also been investigated~\cite{Belanger:2012jn}.

The mixed nature of $\lspone$ not only has implications for the DM relic but also for the Higgs decays. The implications of light (consistent with the LHC limits)  $\lspone,\chipm$  for the Higgs decays $h \rightarrow \gamma \gamma, h\rightarrow \lspi \lspj$ etc.  were already looked into~(see for example~\cite{Belanger:2000tg,Belanger:2001am,Yaguna:2007vm,AlbornozVasquez:2011aa}) even before the Higgs discovery. These had already demonstrated the correlations between the masses of the light EWeakinos and different branching ratios of a light Higgs. The invisible decay width of the $h$ is controlled by the bino-higgsino mixing, apart from the masses of course. Since the same mixing also affects the thermal $\lspone$ relic as well as the detection cross-sections, indeed this  observable forms a very important part of the  light  $\lspone$ phenomenology.
\comment{\footnote{The invisible decays of the $h$ play an important role in the light $\widetilde\nu_R$ DM phenomenology as already mentioned and has been studied, eg. ~\cite{Banerjee:2013fga}.}.} Reference~\cite{Arbey:2012na} discussed comprehensively different aspects of a light $\lspone$ for LHC, SUSY particle spectra and DM detection experiments, on the eve of the Higgs discovery.

The discovery of the Higgs in 2012, removed one big unknown from the situation and thus gave a new direction for the phenomenological studies of the light $\lspone$ case. \comment{The small mass of the Higgs,  absence of light sfermion signals already started pushing the sfermions masses to higher values.} So in addition to the earlier constraints one also had to now impose constraints on the possible parameter space implied by the properties of the observed Higgs~($h_{125}$) and the constraints on sparticle masses put by the non-observation of SUSY at the LHC. Particularly relevant  parameters in the context of a light $\lspone$ are of course the slepton masses, $m_A$ and the couplings of $\lspone$ with $h_{125}, A$ which are controlled by the mixing in the neutralino sector as well as $\tan \beta$. There were many investigations of the light $\lspone$ case in light of the knolwedge of the properties of $h_{125}$ and lack of SUSY signals at the LHC~\cite{Calibbi:2013poa,Arbey:2013aba,Belanger:2013pna,Hagiwara:2013qya}.

The announcement of the Xenon100 results in 2013 
was another game changer, as in one go it removed a part of the parameter space, where a light $\lspone$ could be a good thermal DM. For a nice summary of the situation  which explored the case of $m_{\lspone} < 46$ GeV, see~\cite{Boehm:2013qva}. The discussions of~Refs.\cite{Calibbi:2014lga,Han:2014nba,Cao:2015efs,Hamaguchi:2015rxa} confirmed  that the best way to achieve  a low mass thermal DM in the pMSSM is to 
have a mixed bino-higgsino $\lspone$, the $Z$ and $h_{125}$ exchange (called $Z$ or $h_{125}$ funnel) providing the necessary annihilation, possible light $A$ being ruled out by consideration of the LEP/LHC searches. One should add here that the  bulk region where the slepton co-annihilation works and hence a light, bino-like $\lspone$ is allowed, opens up if one relaxes the requirement of gaugino mass unification, minimal flavour violation and CP conservation from the MSSM~\cite{Fukushima:2014yia}. One should however also note that the last two are generally introduced so as to avoid problems with measurements in  the flavour sector in MSSM and their relaxation is a bit of a tight rope walk in view of the ever more precise measurements in the flavour sector.

In summary, we notice  that in the 19 parameter pMSSM, a light $\lspone$ can be realised in the region $m_{\lspone} \leq 62.5$~GeV and focus on that further.  The lower bound on the chargino mass from measurements at the LEP~($\mchonepm \geq 103.5$~GeV)~\cite{Abbiendi:2003sc} imposes a lower limit on the higgsino and the wino mass parameters~($M_{2},\mu~\gts ~100~{\rm GeV}$) and indeed the $\lspone$ in the above mass region is dominantly a bino. There have been many recent investigations in this context~\cite{Barman:2017swy,Duan:2017ucw,Abdughani:2017dqs,Arbey:2017eos,Pozzo:2018anw}, looking at the viability of light $\lspone$ in light of the latest exclusions in the $\sigma_{SI}$ -- $\mlspone$ plane by the Xenon-1T experiment as well as  the latest EWeakino searches. The situation is summarised in Section \ref{sec:3}. \comment{We see there with the latest results from Xenon 1T experiments the $Z$-funnel region has all but disappeared. The status of the $h_{125}$-funnel region and how it can be tested in all the future experiments,  will be discussed there.} 

As a small digression let us note the following. The large values of the $\lspone$ masses needed in the SUSY models  for a pure higgsino/wino case to explain the relic, have also added to the  questions about  naturalness of the SUSY solution to DM.  In radiative natural SUSY models~\cite{PhysRevLett.109.161802}, the requirement that their EW fine tuning(FT) measure $\Delta_{EW}$ be less than $\sim 30$, implies small values of $\mu$ which naturally indicates  $\lspone$ to be higgsino and `light' ($100 - 300$ GeV). Of course such low mass higgsino can not explain the total observed relic and one is forced to think of a multi component DM. However even that is not 'light' enough by considerations of this article. A generic pMSSM discussion, however, investigating naturalness using the FT measure $\Delta_{EW}$\cite{vanBeekveld:2019tqp,vanBeekveld:2016hug}, does find a cluster of  points around $m_{\lspone} \simeq m_Z/2$ or $\simeq m_{h_{125}}/2$ with low values of FT and a thermal relic density $\leq \Omega^{max}_{obs} h^{2}$. 

Next let us now turn to the case of the NMSSM. The NMSSM,  described by the superpotential of Eq.\ref{eq:wnmssm}, is the simplest extension of the MSSM which was suggested as a mechanism which can explain why $\mu$ is small in a 'natural' way.  The NMSSM Higgs sector is phenomenologically richer than that of MSSM and has an additional CP-even and a CP-odd Higgs state. Among the three CP-even Higgs bosons, $h_{1},~h_{2},~h_{3}$, one is identified with $h_{125}$. In addition, the Higgs sector consists of two CP-odd pseudoscalar Higgs states, $A_{1},~A_{2}$, and two charged Higgs bosons. Along with $\tan\beta$ and $\mu$, additional parameters of the Higgs sector are  $\lambda$,~$\kappa$,~$A_{\lambda}$,~$A_{\kappa}$, where $A_{\lambda}$ and $A_{\kappa}$ are the trilinear soft-breaking parameters~\cite{Ellwanger:2009dp}. The   EWeakino sector also has a new ingredient viz. \textit{singlino}~($\hat{S}$). This results  in $5$ neutralinos and $2$ charginos, and is parameterised by: $M_{1},~M_{2},~\mu,~\tan\beta,~\lambda,~\kappa$. The $5\times 5$ neutralino mass matrix has the following form~(following the notation of Ref.~\cite{PhysRevD.95.115036}):
\begin{small}
\begin{equation}
{\cal M}_{\lspi}=
  \begin{pmatrix}
    M_{1} & 0 & -m_{Z}\sin\theta_{W}\cos\beta & m_{Z}\sin\theta_{W}\sin\beta & 0 \\
    0 & M_{2} & m_{Z}\cos\theta_{W}\cos\beta & -m_{Z}\cos\theta_{W}\sin\beta & 0 \\
    -m_{Z}\sin\theta_{W}\cos\beta & m_{Z}\cos\theta_{W}\cos\beta & 0 & -\mu & -\lambda v\sin\beta \\
    m_{Z}\sin\theta_{W}\sin\beta & -m_{Z}\cos\theta_{W}\sin\beta & -\mu & 0 & -\lambda v\cos\beta \\
    0 & 0 & -\lambda v\sin\beta & -\lambda v\cos\beta & 2\kappa v_{s} \\
    \label{eqn:neut_mass_matrix}
  \end{pmatrix}
\end{equation}
\end{small}
The mass eigenstates in this case are then a  mixture of the $\tilde B, 
\tilde W_3,\tilde h_u^0, \tilde h_d^0$ and the $\tilde S$. The singlino component will then modulate the interactions of the SM particles with the $\lspi, i=1,5$.
Due to the presence of additional structure in the neutralino sector, additional mediators which can be involved in the $\lspone$ annihilation processes and in interactions with nuclei as well as the rich Higgs phenomenology at the colliders and in flavour physics, study of thermal DM in the NMSSM has been a very fertile field of exploration\cite{Ellwanger:2009dp,Belanger:2005kh,Mahmoudi:2010xp}. The case of a light $\lspone$ has been discussed in the literature extensively~\cite{Gunion:2005rw,Ferrer:2006hy,Vasquez:2010ru,Das:2010ww,Cao:2011re,AlbornozVasquez:2011js,Kozaczuk:2013spa,Ellwanger:2013rsa,Cao:2013mqa,Huang:2014cla,Ellwanger:2015axj,Barducci:2015zna,Ellwanger:2016sur,Mou:2017sjf,Ellwanger:2018zxt,Abdallah:2019znp,Wang:2020dtb,Guchait:2020wqn,Barman:2020vzm}\footnote{In addition to these we already discussed the extensions of the NMSSM where a $\widetilde\nu_R$  can be a thermal DM candidate and leads to characteristic LHC signals.}. 

In the NMSSM there exists a  possibility of a light singlet dominated scalar or pseudoscalar  which provides a good annihilation channel for a singlino-higgsino mixed $\lspone$ , thus facilitating the correct relic abundance. Thus one can say that here one can have $h_{1},A_1,Z$ and $h_{125}$ funnel regions where an over dense universe can be avoided. Compatibility with the DD cross-section is obtained because the interactions  of $\lspone$ with quarks are weakened due to the singlino content.The case of light $\lspone$ in the NMSSM was  examined comprehensively in \cite{Vasquez:2010ru}. Due to difference between the DM velocities at the freeze-out and in the galaxies and the flexibility in the couplings of ~$\lspone$, it is possible to have both large or small DD cross-sections for light $\lspone$ in the NMSSM~\cite{Vasquez:2010ru}, for small masses of the mediator $h_1,A_1$. These can escape the constraints from  LEP and the LHC due to their singlet dominated nature.
 
By virtue of the self coupling in the scalar sector, the $h_{125}$ can decay into a pair of $h_1/A_1$ which can then decay into a pair of fermions or a pair of light EWeakinos, thus giving rise to new exotic or invisible decays of the $h_{125}$. Thus these scenarios can be probed through the properties of $h_{125}$, see for example \cite{Huang:2014cla,Wang:2020dtb,Barman:2020vzm}. In addition to these one can also probe them through direct production of the light higgses and the light EWeakinos, at the LHC and future machines~\cite{Mahmoudi:2010xp,Cao:2011re,Ellwanger:2013rsa,Ellwanger:2015axj,Barducci:2015zna,Ellwanger:2016sur,Ellwanger:2018zxt,Abdallah:2019znp,Guchait:2020wqn,Ma:2020mjz,Barman:2020vzm}. This  means that the allowed parameter space of the NMSSM is then constrained by the observed mass and measured signal strengths of the $h_{125}$, current EWeakino searches, searches for light scalars at the LEP and the LHC, the latter through the production in $h_{125}$ decay. Of special importance are also the constraints from flavour physics as was pointed in~\cite{Gunion:2005rw}.  Further, for these light $\lspone$'s, the Indirect DM detection experiments, looking at the radio emission in the Milky Way and in galaxy clusters, gamma rays in the dwarf spheroidal (dSph) galaxies and also the antiprotons in the milky way~\cite{Ferrer:2006hy,AlbornozVasquez:2011js,Huang:2014cla}, can constrain the NMSSM parameter space very effectively. The constraints from the Xenon-1T experiment have again played an important role in constraining the parameter space. In fact, the direct and indirect detection experiments are pretty complementary in this  case~\cite{AlbornozVasquez:2011js}. Section~\ref{sec:4} contains a detailed discussion of the current status of a light $\lspone$ in NMSSM and discovery prospects at the LHC (HL/HE) , the future DM detection experiments and the $e^+e^-$ colliders: the Higgs and the $B$-factories.

\section{Light $\lspone$ DM in pMSSM}
\label{sec:3}

The impetus of this section is the pMSSM scenario with a light neutralino DM and with parameters defined at the electroweak scale. Let us first discuss the case where $\lspone$ is a thermal relic~($\Omega_{\lspone} h^{2} \leq \Omega_{obs}^{max}h^{2}$). Recent works in this direction~\cite{Barman:2017swy,Pozzo:2018anw}~(and the references therein) have explored the impact of current limits from collider, astrophysical and cosmological measurements. 
The analysis in reference~\cite{Barman:2017swy} considered the following range of input parameters:
\begin{eqnarray}
1~{\rm GeV}~<~&& M_{1}~<~100~ {\rm GeV}, \quad 90~{\rm GeV}~<~ M_{2}~<~3~ {\rm TeV},\nonumber\\
 \quad  1~<~&&\tan{\beta}~<~55,\quad 70~{\rm GeV}~<~ \mu~<~3~ {\rm TeV}, \nonumber \\
 \quad  800~{\rm GeV}~<~&& M_{\tilde{Q}_{3l}} < 10~{\rm TeV}, \quad 800~{\rm GeV}~<~ M_{\tilde{t}_{R}} < 10~{\rm TeV}, \nonumber \\
 \quad && 800~{\rm GeV}~<~ M_{\tilde{b}_{R}} < 10~{\rm TeV}, \nonumber \\ 
\quad 2~{\rm TeV}~<~&&M_{3}~<~5~ {\rm TeV}, \quad -10~{\rm TeV}~<~ A_{t}~<~10~ {\rm TeV}
\label{Parameter_space}
\end{eqnarray}
\noindent The mass of the first and second generation squarks and the sleptons were fixed at $3~{\rm TeV}$, while both $A_{b}$ and $A_{\tau}$ were taken to be $0$. The pseudoscalar mass was fixed at $1~{\rm TeV}$ in order to decouple its effect from DM phenomenology and the lighter CP-even Higgs boson was identified with the $125~{\rm GeV}$ Higgs boson of the Standard Model~($h_{125}$).

\begin{figure}[!htb]
\begin{center}
\resizebox{0.60\columnwidth}{!}{
\includegraphics{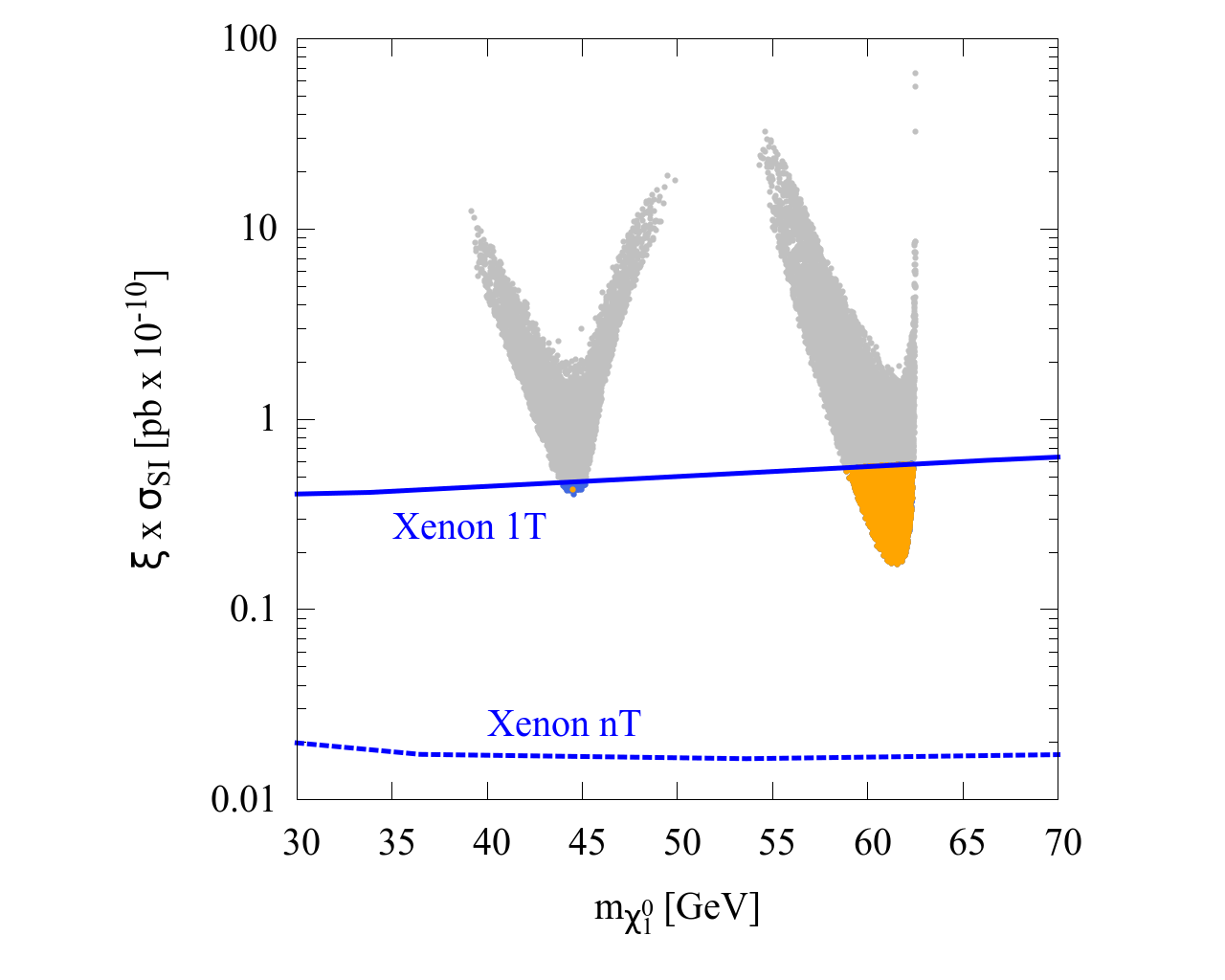}}
\caption{SI DM-nucleon cross-section vs $\mlspone$ for all points allowed by the collider and relic density constraints~(modified from Ref.~\cite{Barman:2017swy}). The reach of various DD experiments is shown by lines labelled accordingly.} 
\label{fig:MSSM_thermal_current}
\end{center}
\end{figure}

The parameter space considered in reference~\cite{Barman:2017swy}~(Eqn.~\ref{Parameter_space}) was restricted to the $\mu > 0$ regime where the DD limits on $\sigma_{SI}$ are more sensitive than their spin-dependent counterparts. The current limits~(projected reach) on $\sigma_{SI}$ from Xenon-1T~\cite{Aprile:2018dbl}~(Xenon-nT~\cite{Aprile:2015uzo}) at $90\%$ CL are illustrated as a solid~(dashed) blue line in Fig.~\ref{fig:MSSM_thermal_current}. The vertical axis in Fig.~\ref{fig:MSSM_thermal_current} represents $\sigma_{SI}$ rescaled with $\xi$, defined as the ratio of the predicted relic density of $\lspone$~($\Omega_{\lspone}h^{2}$) to $\Omega_{obs}^{max}h^{2}$
\begin{equation}
\xi = \frac{\Omega_{\lspone}h^{2}}{0.122}
\end{equation} 
Note that in the framework of the pMSSM considered here, the Higgs signal strength constraints impose an indirect upper limit on the Higgs to invisible branching fraction~($\lesssim 10\%$), which is stronger than the current direct limit. 
The grey points in Fig.~\ref{fig:MSSM_thermal_current} are excluded by the latest upper limits from Xenon-1T. The excluded points are typically those which have a relatively larger higgsino component in $\lspone$. The bounds from Xenon-1T excludes almost all the points in the $Z$ funnel region and a significant fraction of points in the $h_{125}$ funnel region. Note that the constraints from Xenon-1T exclude points which were otherwise allowed by the Higgs signal strength constraints. 

Let us next discuss the impact of current limits from direct EWeakino searches. For $\mlspone \leq 62.5~{\rm GeV}$, the current exclusion limit at $95\%$ CL on pure higgsinos and pure winos from a combination of direct EWeakino searches in the $WZ$ mediated $3l+\met$ channel
, $2l+\met$ channel 
 and the $Wh_{125}$ mediated $1l+2b+\met$ channel
, performed using the LHC Run-II data~($\mathcal{L}\sim 35~{\rm fb^{-1}}$), is roughly $390~{\rm GeV}$~\cite{Pozzo:2018anw} and $650~{\rm GeV}$~\cite{Sirunyan:2018ubx}, respectively. 
We impose these limits conservatively by choosing only parameter space  points which have  higgsino-~(wino-) dominated EWeakinos~(with higgsino~(wino) admixture $> 90\%$) with mass less than $390~{\rm GeV}$ ($650~{\rm GeV}$). In Fig.~\ref{fig:MSSM_thermal_current},
the blue coloured points are excluded upon the application of these direct search limits, while the orange coloured points represent the currently allowed parameter space\footnote{Similar results have also been reported in reference~\cite{Pozzo:2018anw}.}. Only one point in the $Z$ funnel region appears to avoid  the current EWeakino constraints. A closer observation reveals that this particular point has $\mu \sim 240~{\rm GeV}$ and $M_{2} \sim 470~{\rm GeV}$
resulting in higgsino-dominated $\lsptwo$, $\lspthree$ and $\chonepm$ with a mass smaller than $390~{\rm GeV}$ and wino-dominated $\lspfour$ and $\chtwopm$ with a mass smaller than $650~{\rm GeV}$. However, the amount of higgsino admixture in $\lsptwo$ and $\chonepm$ is around $88\%$
and the amount of wino admixture in $\lspfour$ and $\chtwopm$ is around $89\%$, thus, falling only marginally outside the conservative interpretation of the current reach of direct higgsino and wino searches. Fig.~\ref{fig:MSSM_thermal_current} also shows that the entire region of currently allowed parameter space in the thermal scenario falls within the projected reach of the Xenon-nT.

\begin{figure}[!htb]
\begin{center}
\resizebox{1.00\columnwidth}{!}{
\includegraphics{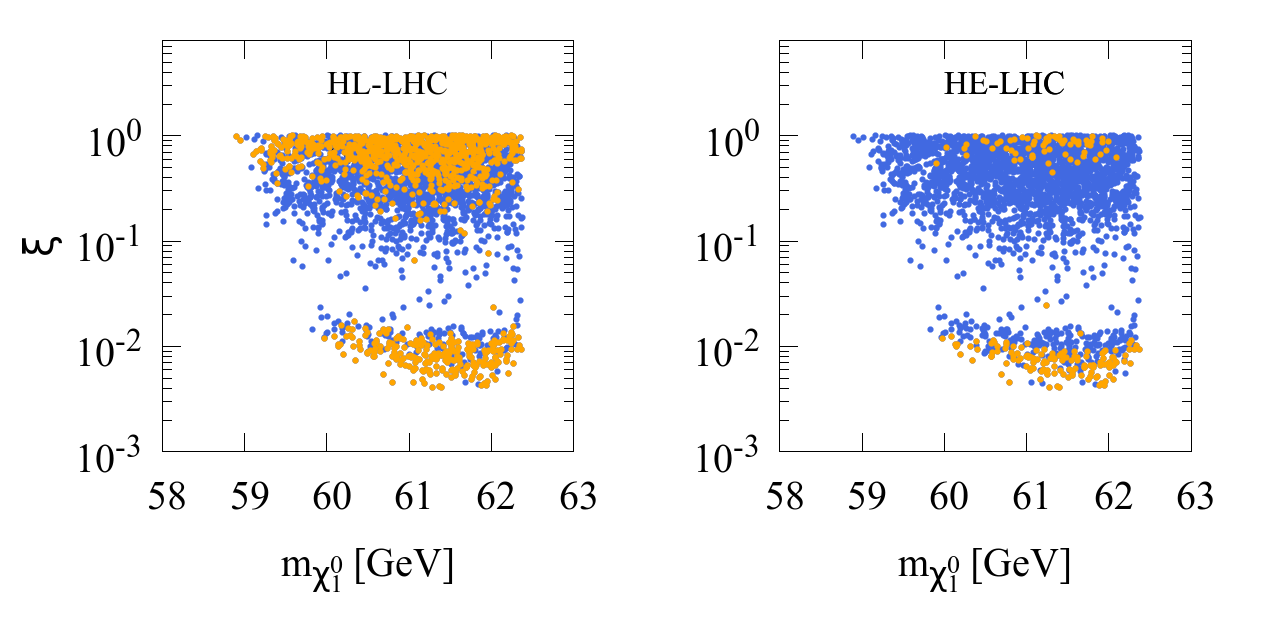}}
\caption{$\xi$ vs $\mlspone$ for the allowed parameter space points. Blue~(Orange) coloured points fall within~(outside) the projected discovery reach of direct searches in the $3l+\met$ channel at the HL-LHC~(left panel) and the HE-LHC~(right panel)~(from Ref.~\cite{inprepmssm:2020}).}
\label{fig:MSSM_thermal_WZ_HL-LHC}
\end{center}
\end{figure}

An ongoing work~\cite{inprepmssm:2020} has analysed the projected capability of the HL-LHC and the HE-LHC ( $\sqrt{s}=27~{\rm TeV}$, $\mathcal{L}=15~{\rm ab^{-1}}$ ) to probe the currently allowed parameter space via direct EWeakino searches in the $WZ$ and $Wh_{125}$ mediated $3l+\met$ final state. The analysis in reference~\cite{inprepmssm:2020} utilises the signal regions and efficiency grids obtained in reference~\cite{Barman:2020vzm}\footnote{Details on the translation scheme can be found in reference~\cite{Barman:2020vzm}}. 
The projected reach of the HL-LHC (HE-LHC) is shown in Fig.~\ref{fig:MSSM_thermal_WZ_HL-LHC}~left (right) panel  for the currently allowed parameter space in the $h_{125}$ funnel region. The currently allowed points are mostly concentrated in the $\xi \lesssim 0.02$ and $\xi \gtrsim 0.1$ regions which is an implication of the current limits from direct EWeakino searches. The points in the $\xi \lesssim 0.02$ region have $\mu \lesssim 150~{\rm GeV}$ and $M_{2} \gtrsim 650~{\rm GeV}$. As a result, the amount of bino admixture in $\lsptwo$ and $\lspthree$ increases which in turn leads to a reduction in the higgsino composition to values below $ 90\%$. These points, therefore, survive the current limits from direct higgsino searches. Since, $M_{2}$ is also larger than $\sim 650~{\rm GeV}$, the wino-like $\lspfour$ and $\chtwopm$ evades the constraints from direct wino searches. The region with $\xi \gtrsim 0.1$ is mostly populated by points where $M_{2} \gtrsim 250~{\rm GeV}$ and $\mu \gtrsim 400~{\rm GeV}$. Consequently, the heavier neutralinos and the charginos are either wino-higgsino mixed states or outside the current limits from direct EWeakino searches. Furthermore, the large mass difference between $M_{1}$ and $\mu$ results in a relatively smaller $h_{125}\lspone\lspone$ coupling leading to higher values of $\xi$. The intermediate $\xi$ region~(0.02 $\lesssim \xi \lesssim$ 0.1) with smaller density of points have $150~{\rm GeV} \lesssim \mu \lesssim 400~{\rm GeV}$ and $M_{2} \lesssim 500~{\rm GeV}$. The points with $M_{2} \gtrsim 500~{\rm GeV}$ get excluded by the current limits from direct higgsino searches since the amount of higgsino content in $\lsptwo/\lspthree$ and $\chonepm$ increases above $90\%$.  
The only allowed point in the $Z$ funnel region also falls within the projected discovery reach of direct EWeakino searches at the HL-LHC. 

In Fig.~\ref{fig:MSSM_thermal_WZ_HL-LHC}, the orange~(blue) coloured points are outside~(within) the projected discovery reach. HE-LHC displays a much larger discovery reach compared to the HL-LHC. The orange points close to $\xi \sim 10^{-2}$ have a typically large $M_{2}$ ($\gtrsim 1~{\rm TeV}$) and a small $\mu$~($\lesssim 150~{\rm GeV}$). As a result, the $\lsptwo,~\lspthree$ and $\chonepm$ in these parameter space points have a dominant higgsino composition, and $pp \to \lsptwo\chonepm +\lspthree\chonepm$ are the dominant chargino-neutralino pair production modes. However, due to smaller value of $\mu$, the mass difference between $\lsptwo/\lspthree/\chonepm$ and $\lspone$ is either small or very close to the $W$/$Z$/$h_{125}$ mass resulting in suppressed signal efficiencies in both $WZ$ and $Wh_{125}$ mediated channels. 
As a result, despite having large production rates, these points result in a very small or zero signal significance. The orange points close to $\xi \sim 1$ typically have large $M_{2}$ and $\mu$ and falls outside the projected discovery reach due to small production cross-section. 

There is also substantial motivation to consider the scenario, $\Omega_{\lspone} h^{2} > \Omega_{obs}^{max}h^{2}$, where non-standard mechanisms enable the production of the observed relic density~(see Sec.~\ref{subsec:2.1}). Considering the parameter space shown in equation~\ref{Parameter_space}, the work in reference~\cite{Barman:2017swy} shows that such a scenario can lead to phenomenological features which are distinct from the predictions of the thermal relic scenario. The lower limit on $\mlspone$ is lifted in the non-standard scenario and allowed points are obtained with $\mlspone$ as small as $\lesssim 1~{\rm GeV}$ up to $62.5~{\rm GeV}$. Relaxing the relic density constraint allows very small values of $h_{125}\lspone\lspone$ couplings since an efficient annihilation is no longer required. As a result, a wide range of $Br(h_{125} \to \lspone\lspone)$ is observed which can attain values as small as $\sim 10^{-6}$. 

\begin{figure}[!htb]
\begin{center}
\resizebox{1.00\columnwidth}{!}{
\includegraphics{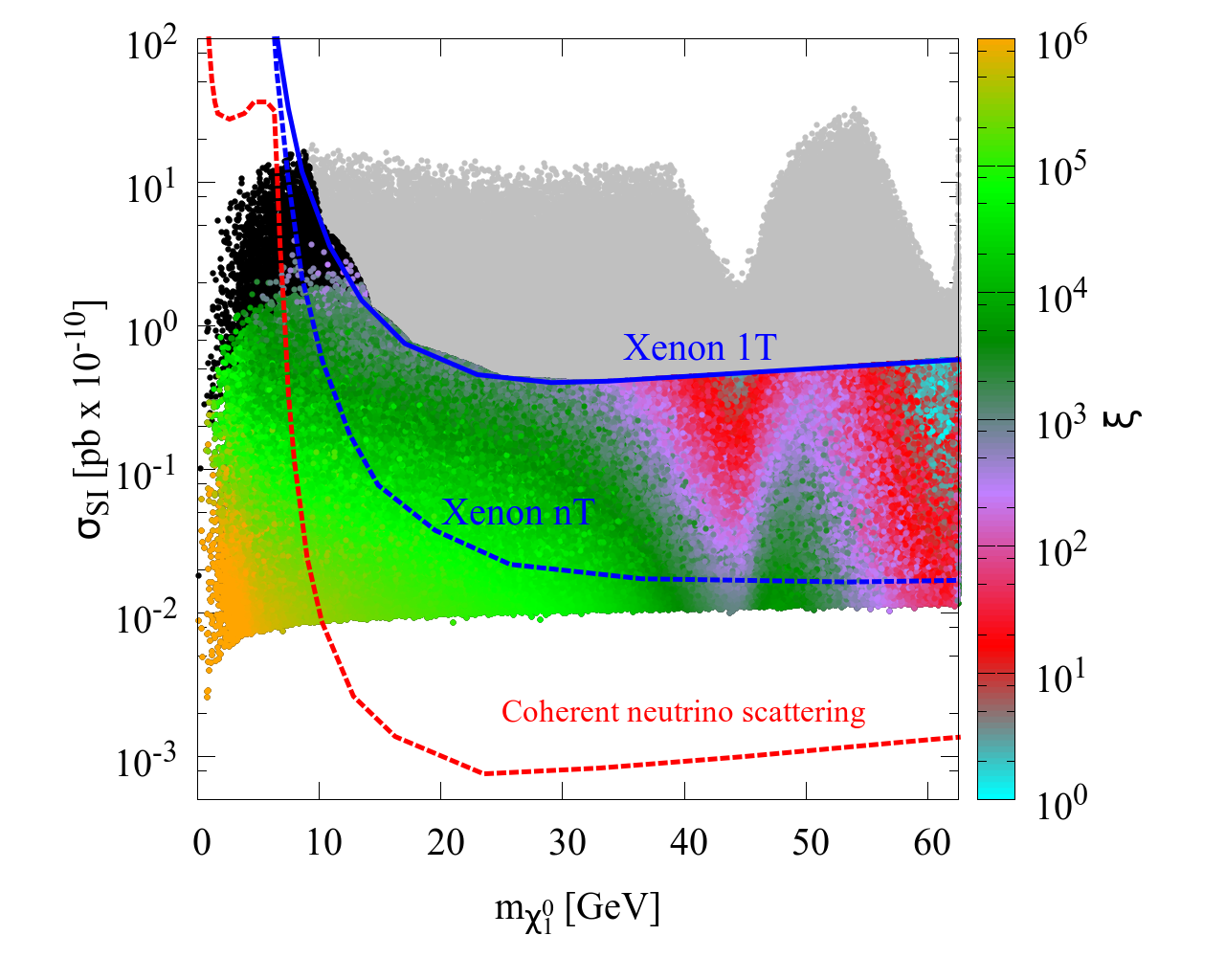}\includegraphics{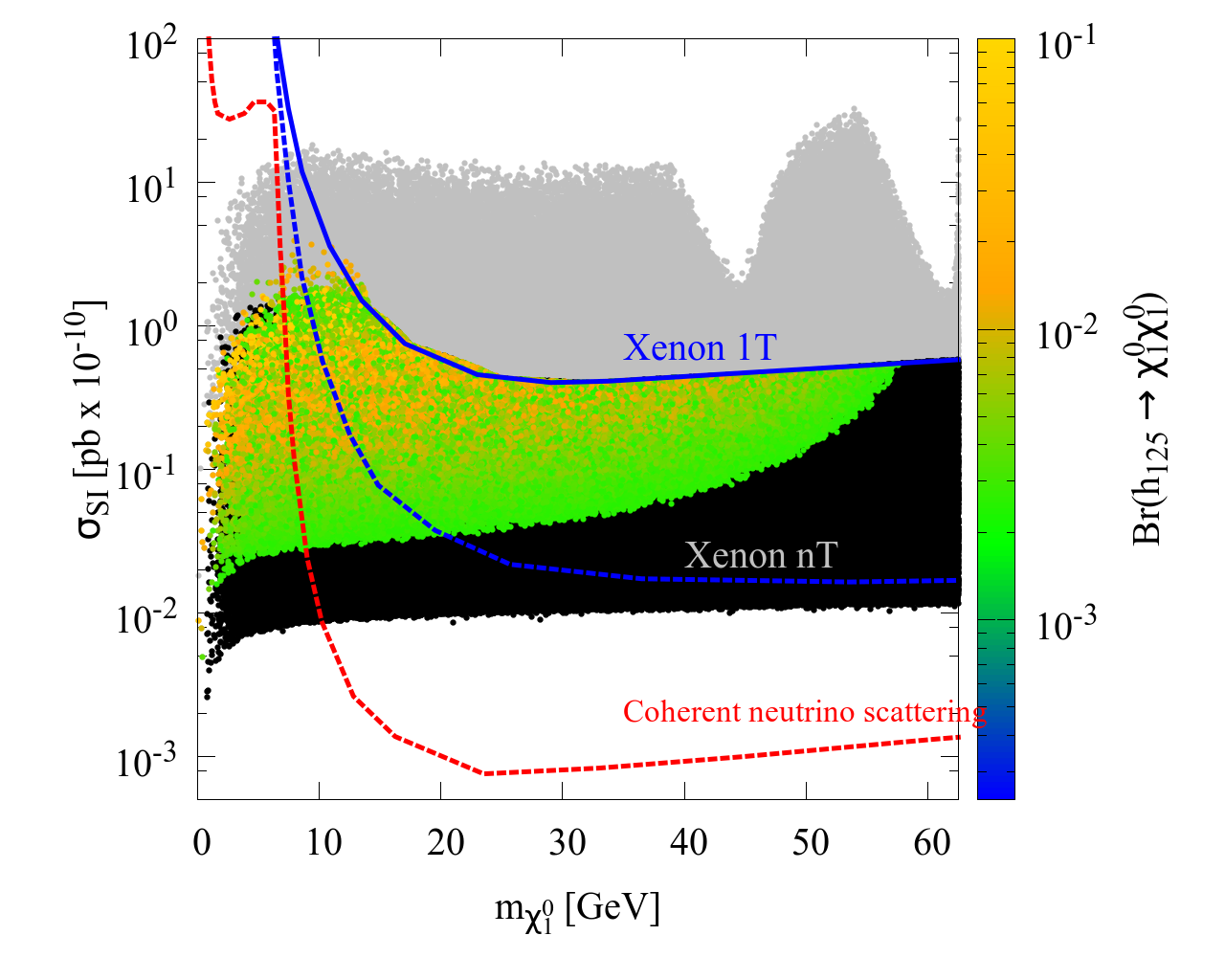}}
\caption{Points allowed by the collider and relic density constraints~($\Omega_{\lspone} h^{2} > 0.122$) are illustrated in the $\sigma_{SI}$ vs $\mlspone$ plane. \textbf{Left panel}~(from Ref.~\cite{inprepmssm:2020}): Grey points are excluded by the current Xenon-1T limits and the black points are excluded by the current limits from direct EWeakino searches. The colour palette shows the variation in $\xi$. \textbf{Right panel}~(modified from Ref.~\cite{Barman:2017swy}): Grey points are excluded by the current Xenon-1T constraints and the current LHC limits. Black coloured points have $Br(h_{125} \to invisible) < 0.24\%$. The colour palette shows the variation in $Br(h_{125} \to invisible)$. The reach of various DD experiments is shown by lines labelled accordingly.
}
\label{fig:MSSM_nonthermal_current}
\end{center}
\end{figure}

The points corresponding to $\Omega_{\lspone}h^{2} > 0.122$
and allowed by the collider constraints are illustrated in the $\sigma_{SI}$-$\mlspone$ plane in Fig.~\ref{fig:MSSM_nonthermal_current}.  A significantly large population of points are allowed by the current DD constraints over the entire $\mlspone$ range contrary to the thermal scenario. The black coloured points in the left panel of Fig.~\ref{fig:MSSM_nonthermal_current} are excluded by the combined limits from direct EWeakino searches~\cite{Pozzo:2018anw,Sirunyan:2018ubx}. The colour palette is used to illustrate the variation in $\xi$ and the coloured points represent the currently allowed parameter points. The funnel regions correspond to smaller values of $\xi$ due to annihilation via resonance and $\xi$ attains larger values as one moves towards smaller $\mlspone$ values.

In Fig.~\ref{fig:MSSM_nonthermal_current}~(right panel), the complementarity between the DD cross-sections and Higgs to invisible branching is highlighted. The grey points are excluded by the current limits from Xenon-1T and the direct EWeakino searches. 
Currently allowed points with $Br(h_{125} \to \lspone\lspone) < 0.24\%$ are shown in black colour\footnote{The CEPC is projected to be capable of probing the Higgs to invisible branching fraction as small as $\sim 0.24\%$~\cite{CEPCStudyGroup:2018ghi}.} and therefore, are outside the projected Higgs to invisible measurement capability of the CEPC. The points which can be probed at the CEPC are illustrated with a colour palette representing the variation in $Br(h_{125} \to \lspone\lspone)$. Fig.~\ref{fig:MSSM_nonthermal_current}~(right panel) shows that a large fraction of points which are outside the projected reach of Xenon-nT can be probed at the CEPC through measurements of the invisible branching fraction of the Higgs. Furthermore, all the points illustrated in Fig.~\ref{fig:MSSM_nonthermal_current} fall outside the current limits as well as the projected reach~(at $90\%$ CL) of the SuperCDMS experiment~\cite{Agnese:2014aze,Agnese:2016cpb} which aims at directly detecting the low mass WIMPs~($\lesssim 10~{\rm GeV}$). For $m_{\lspone} \sim 8~{\rm GeV}$, the current upper limit from SuperCDMS exclude $\sigma_{SI} \gtrsim 10^{-42}~{\rm cm^{-2}}$ while its projected sensitivity reaches up to $\sigma_{SI} \sim 10^{-44}~{\rm cm^{-2}}$. The current upper limits and the future projections from SuperCDMS have not been illustrated in Fig.~\ref{fig:MSSM_nonthermal_current} since they  fall outside the range of the $y$-axis.

\begin{figure}[!htb]
\begin{center}
\resizebox{1.00\columnwidth}{!}{
\includegraphics{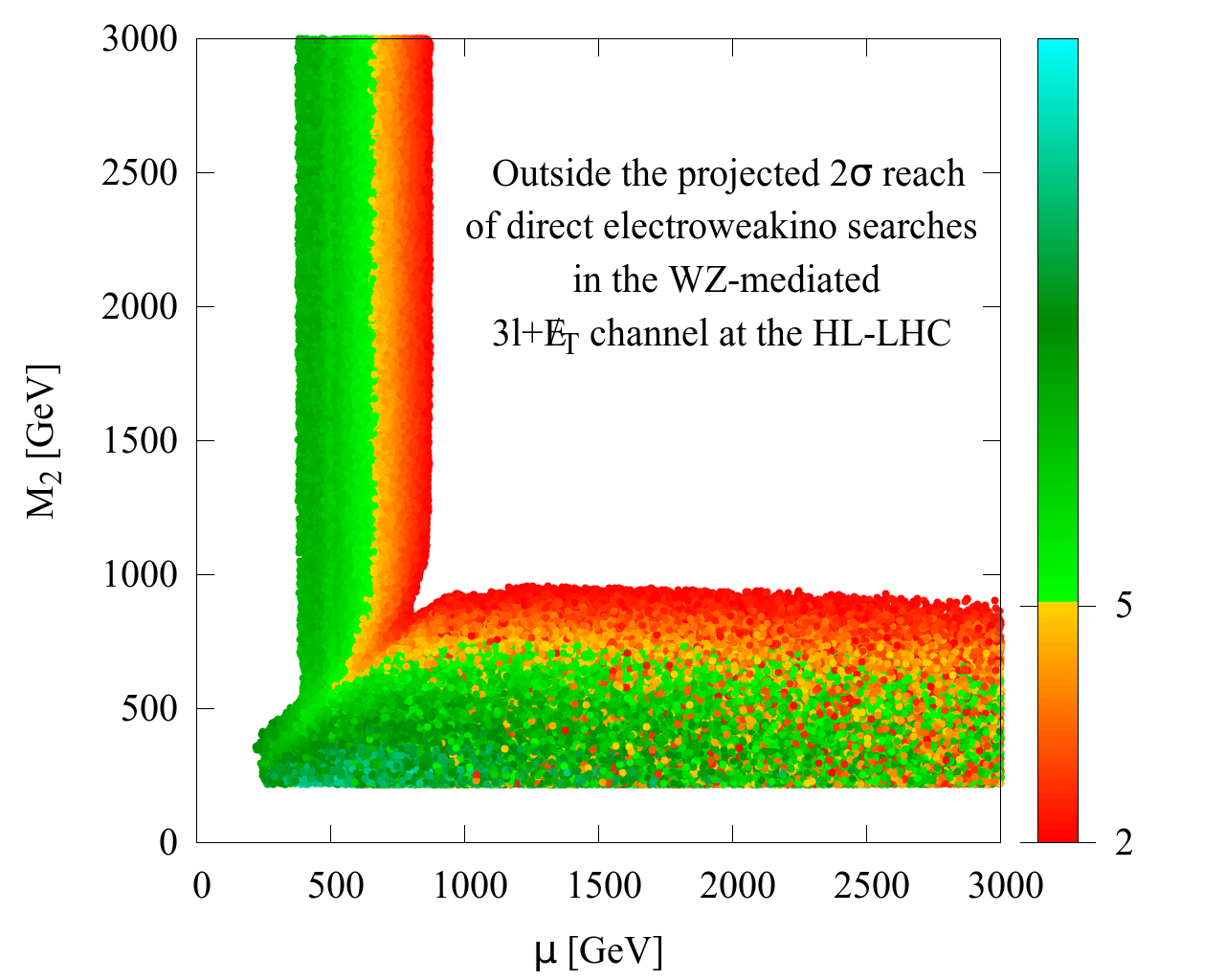}\includegraphics{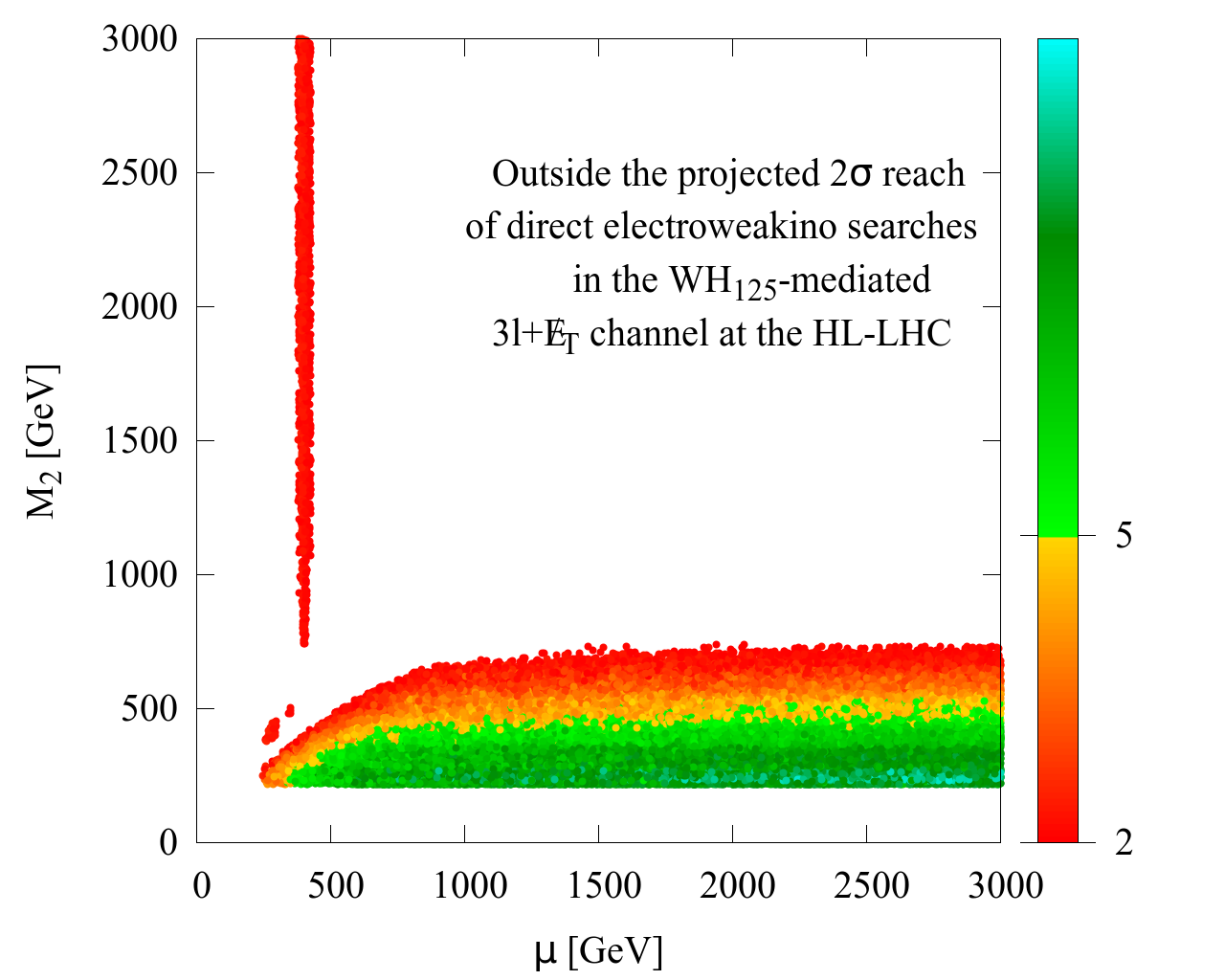}}
\caption{$M_{2}$ vs $\mu$ for those allowed parameter space points which are within the projected exclusion reach of direct EWeakino searches in the $WZ$ mediated~(left panel) and $Wh_{125}$ mediated~(right panel) $3l+\met$ channel at the HL-LHC. The colour palette shows the variation of signal significance in the respective channels~(from Ref.~\cite{inprepmssm:2020}).}
\label{fig:MSSM_nonthermal_HL-LHC}
\end{center}
\end{figure}

The projected capability of the HL-LHC to probe the allowed parameter space via direct EWeakino searches in the $WZ$ and $Wh_{125}$ mediated $3l+\met$ channels is shown in the left and right panels of Fig.~\ref{fig:MSSM_nonthermal_HL-LHC}~(from \cite{inprepmssm:2020}), respectively. The analysis in \cite{inprepmssm:2020} utilises the projection contours derived in \cite{Barman:2020vzm}. Direct EWeakino searches in the $WZ$ mediated channel are able to probe $M_{2}$ up to $\sim 700~{\rm GeV}$ with discovery reach over the entire scanned range of $\mu$. Similarly, the discovery region extends up to $\mu \sim 650~{\rm GeV}$ for all values of $M_{2}$. Note that the $WZ$ mediated channel displays a stronger reach in the $M_{2} > \mu$ while the $Wh_{125}$ mediated channel shows greater sensitivity in the $\mu > M_{2}$ region, thus both  channels are  complementary as clearly seen in Fig.~\ref{fig:MSSM_nonthermal_HL-LHC}.
In the $\mu \lesssim 700~{\rm GeV}$ region, as one further move towards smaller $\mu$ values, the amount of higgsino admixture in $\lsptwo$ and $\chonepm$ increases. This increase in the higgsino admixture leads to an increase in $Br(\lsptwo \to Z\lspone)$ while causing a relative decrease in $Br(\lsptwo \to h_{125}\lspone)$. This results in a smaller signal yield in the $Wh_{125}$ mediated $3l+\met$ channel, thereby, falling outside its projected exclusion reach. The study in \cite{Barman:2020vzm} shows that the direct EWeakino searches at the HE-LHC would be able to probe $\mu$~($M_{2}$) up to $\sim 1050~{\rm GeV}$ over the entire scanned range of $M_{2}$~($\mu$) with discovery reach.

\begin{figure}[!htb]
\begin{center}
\resizebox{1.00\columnwidth}{!}{
\includegraphics{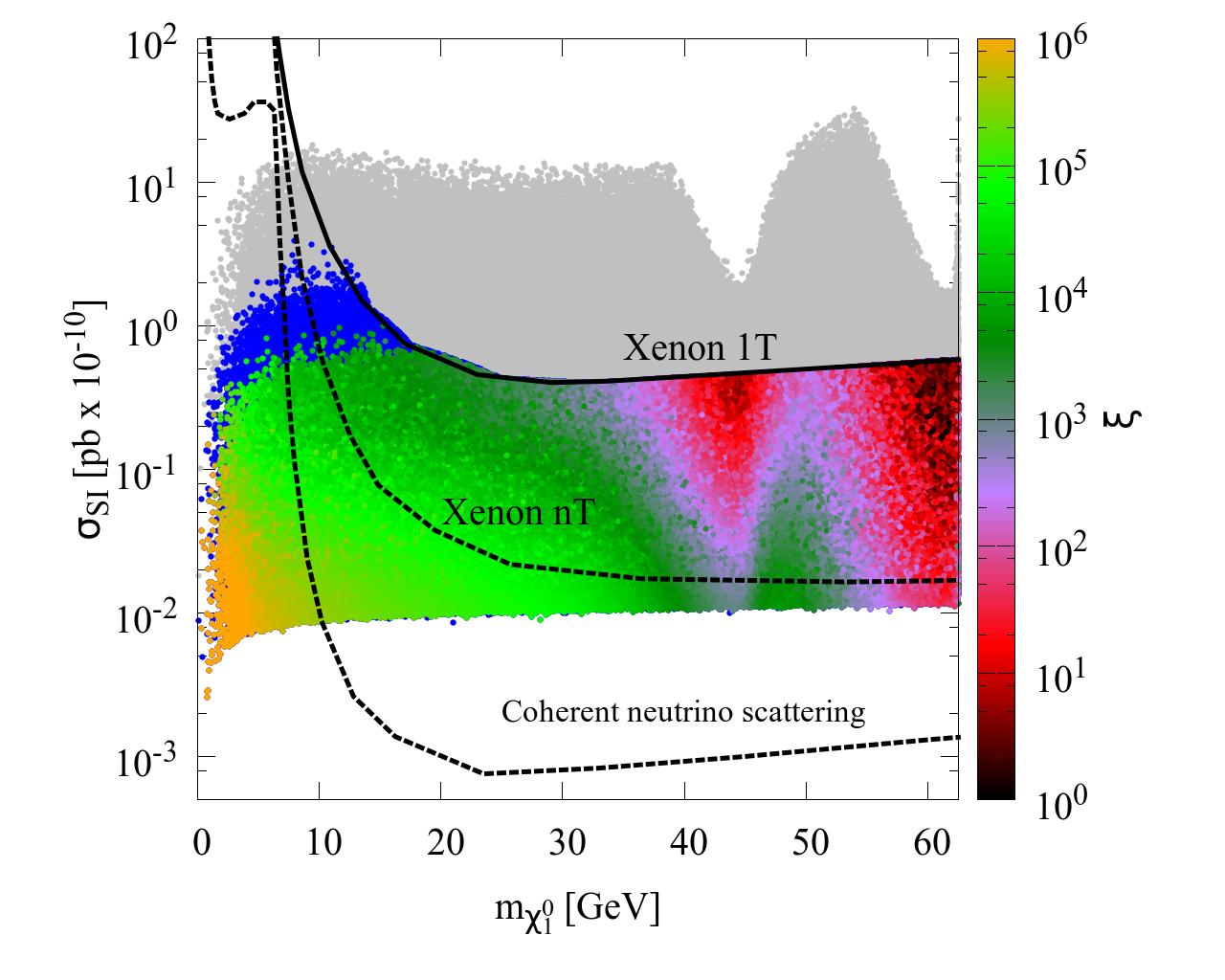}\includegraphics{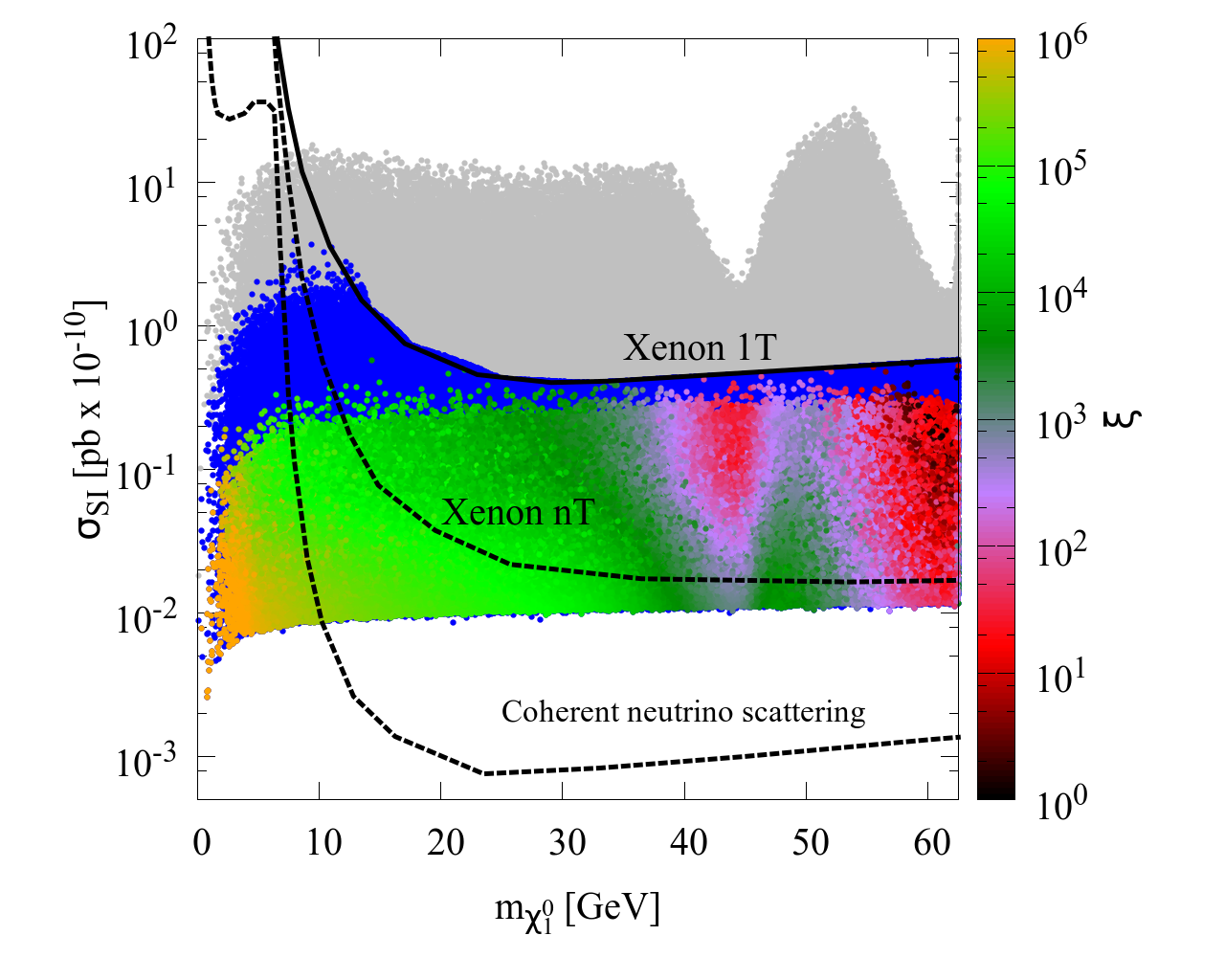}}
\caption{SI DM-nucleon scattering cross-section vs $\mlspone$ for all points allowed by the collider constraints and with $\Omega h^{2} > 0.122$. Grey coloured points are excluded by the current limits from Xenon-1T and the combined limits from direct EWeakino searches. The overlapping colour palette shows the variation in $\xi$ and represents those currently allowed parameter space points which fall outside the projected discovery reach of direct EWeakino searches at the HL-LHC~(left panel) and the HE-LHC~(right panel). The underlying blue coloured points result in a signal significance of $> 5\sigma$ in the $WZ$ and/or $Wh_{125}$ mediated $3l+\met$ search channels at the HL-LHC~(left panel) and the HE-LHC~(right panel)~(from Ref.~\cite{inprepmssm:2020}). The reach of various DD experiments is shown by lines labelled accordingly.}
\label{fig:MSSM_nonthermal_Ewino}
\end{center}
\end{figure}

Before concluding this section, let us take a look at the complementarity between future DD experiments and the direct EWeakino searches at the future LHC in Fig.~\ref{fig:MSSM_nonthermal_Ewino}~(from Ref.~\cite{inprepmssm:2020}). The grey points in the left and right panels of Fig.~\ref{fig:MSSM_nonthermal_Ewino} are excluded by the current constraints. The blue points in the left and right panels fall within the projected discovery reach of direct EWeakino searches in the $3l+\met$ final state at the HL-LHC and the HE-LHC, respectively. The currently allowed parameter space points which also fall outside the projected discovery reach of direct EWeakino searches at the HL-LHC~(left panel) and the HE-LHC~(right panel) are illustrated through an overlapping colour palette showing variation of $\xi$. Note that the blue coloured region extend underneath the coloured points. This implies that the HL-LHC and the HE-LHC will be able to probe even such points which fall below the projected sensitivity of Xenon-nT over the entire LSP mass range via direct EWeakino searches. This result is particularly important for the $\mlspone \lesssim 10~{\rm GeV}$ region where the DD experiments start losing sensitivity.

\section{Light $\lspone$ DM in NMSSM}
\label{sec:4}
The NMSSM allows the possibility of much lighter neutralinos with $\mlspone \sim 1~{\rm GeV}$ while still satisfying the current constraints and providing a substantial component of relic DM~($\Omega_{\lspone}h^{2} \leq \Omega_{obs}^{max} h^{2}$). By virtue of the chargino mass constraint as discussed in Section~\ref{subsec:susy_DM}, the $\lspone$ has to be either bino-dominated or singlino-dominated in the $\mlspone \leq 62.5~{\rm GeV}$ region. In order to satisfy the upper bound on the relic density, the bino or singlino-like $\lspone$ has to undergo co-annihilation or annihilation via resonance. The work in \cite{Barman:2020vzm} focuses on the second possibility and  considers the parameter space with $M_{1}$ fixed at $2~{\rm TeV}$ in order to study beyond-the-MSSM-like region of parameter space, thus, allowing the possibility of an exclusively singlino-like LSP with small admixtures from higgsinos and winos. The added advantage of a singlino-like $\lspone$ is that it can couple with a singlet-like pseudoscalar or scalar Higgs state even in the absence of any higgsino admixture with the coupling being proportional to $\sim \kappa N_{15}^{2}$ where $N_{15}$ refers to the singlino component in $\lspone$. The $Z$ and the Higgs boson provide resonance enhancements in the $Z$ and $h_{125}$ funnel regions. However, at LSP masses below $\sim m_{Z}/2$, efficient annihilation can be realised only in the presence of light singlet-like~(singlet fraction $\gtrsim 90\%$) scalar or pseudoscalar Higgs bosons at roughly twice the mass of the LSP neutralino, $m_{A_1,h_{1}} \sim 2~\mlspone$ as shown in  \cite{Barman:2020vzm}. Keeping these correlations in mind, the following region of parameter space is considered in reference~\cite{Barman:2020vzm}:

\begin{eqnarray}
0.01 < \lambda < 0.7,~10^{-5} < \kappa < 0.05,~3<\tan\beta < 40   \nonumber \\
100~{\rm GeV}< \mu < 1~{\rm TeV}, ~1.5~{\rm TeV}< M_{3} < 10~{\rm TeV} \nonumber \\
2~{\rm TeV} < A_{\lambda}<10.5~{\rm TeV},~-150~{\rm GeV} <A_{\kappa} <100~{\rm GeV} \\
M_{1} = 2~{\rm TeV},~70~{\rm GeV} < M_{2} <2~{\rm TeV},~  A_{t} = 2~{\rm TeV},~A_{b,\tilde{\tau}} = 0 \nonumber 
\label{Eqn:Paramater_scan_NMSSM}
\end{eqnarray}

The third generation squark mass parameters were fixed at $2~{\rm TeV}$ while the first and second generation quark and slepton masses, and the third generation slepton masses were fixed at $3~{\rm TeV}$. 

In the low DM mass region, $\mlspone \lesssim 10~{\rm GeV}$\footnote{$m_{\lspone}$ has been restricted to values larger than $1~{\rm GeV}$.}, the enhancement in the $\lspone$-$\lspone$ annihilation cross-section is realised by the virtue of $h_{1}/A_{1}$ having very narrow widths~($\Gamma/m_{h_{1}/A_{1}} \sim 10^{-7}$-$~10^{-9}$) along with very small couplings due to their singlet nature. This results in a strong velocity dependence of the $\lspone\lspone$ annihilation cross-section. In the early universe, the thermal energy of the DM will lead to the required enhancement in the annihilation cross-section leading to $\Omega_{\lspone}h^{2} \leq 0.122$ and at times $\xi$ closer to $1$. However, this enhancement will be absent at lower galactic velocities, thus allowing the parameter space points to escape the indirect detection constraints from FermiLAT~\cite{Fermi-LAT:2016uux}. There can be exceptions when $\mlspone$ is very near $m_{h_{1}/A_{1}}/2$ or slightly above. In such cases, the resonant enhancement can be realised even at lower DM galactic velocities~($v \sim 10^{-3} c$) and only the tail of the resonance contributes at the higher velocities in the early universe~($v \sim 0.3c$). Within this Breit Wigner enhancement scenario, the DM annihilation cross-section in the galaxy can be larger or comparable with its early universe counterpart~(see for example \cite{Ibe:2008ye,AlbornozVasquez:2011js}). The work in reference~\cite{Barman:2020vzm} observed a few parameter space points which fall within this fine-tuned category and they were excluded by the current constraints from FermiLAT~\cite{Fermi-LAT:2016uux}.

\begin{figure}[!htb]
\begin{center}
\resizebox{1.00\columnwidth}{!}{
\includegraphics{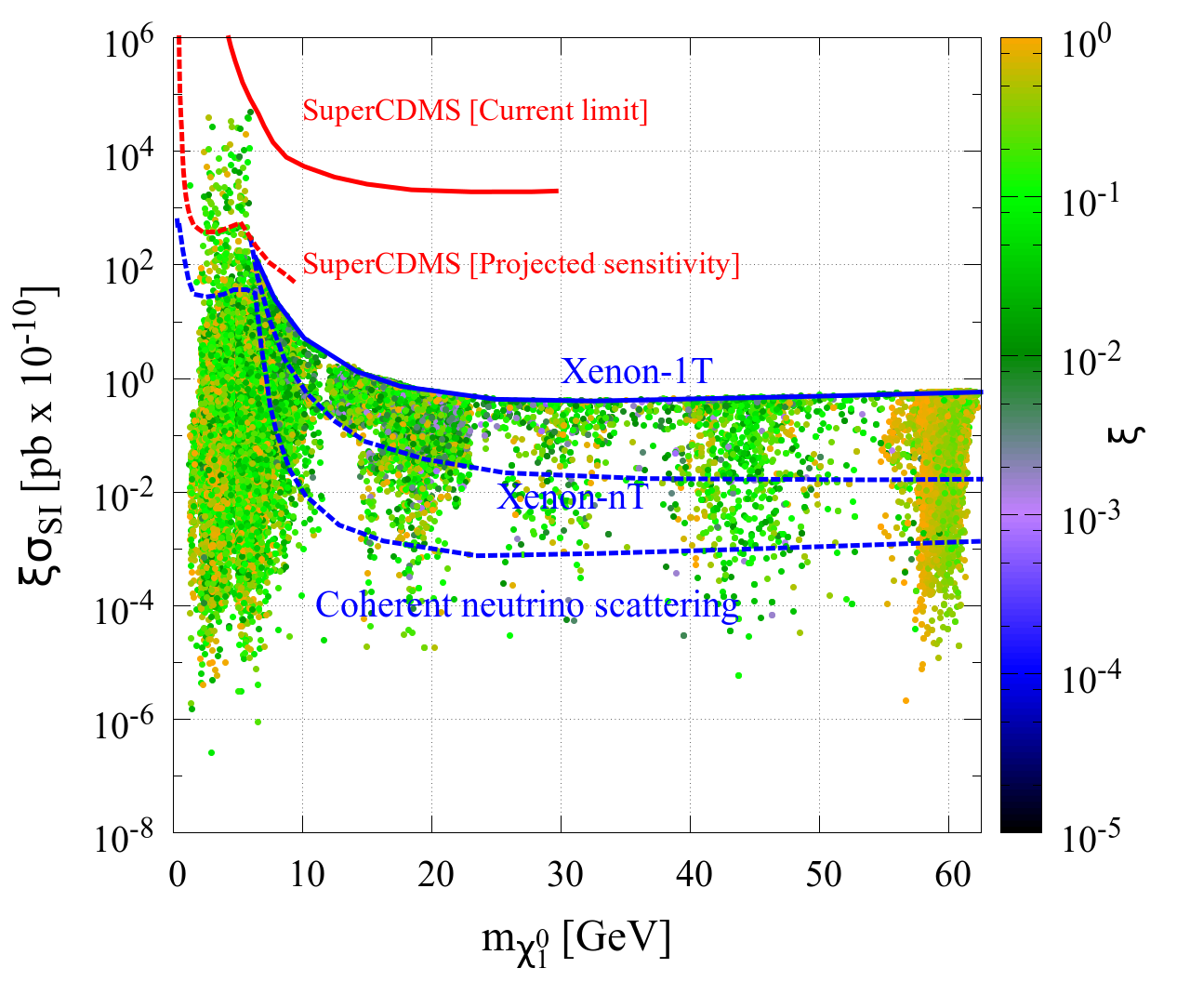}\includegraphics{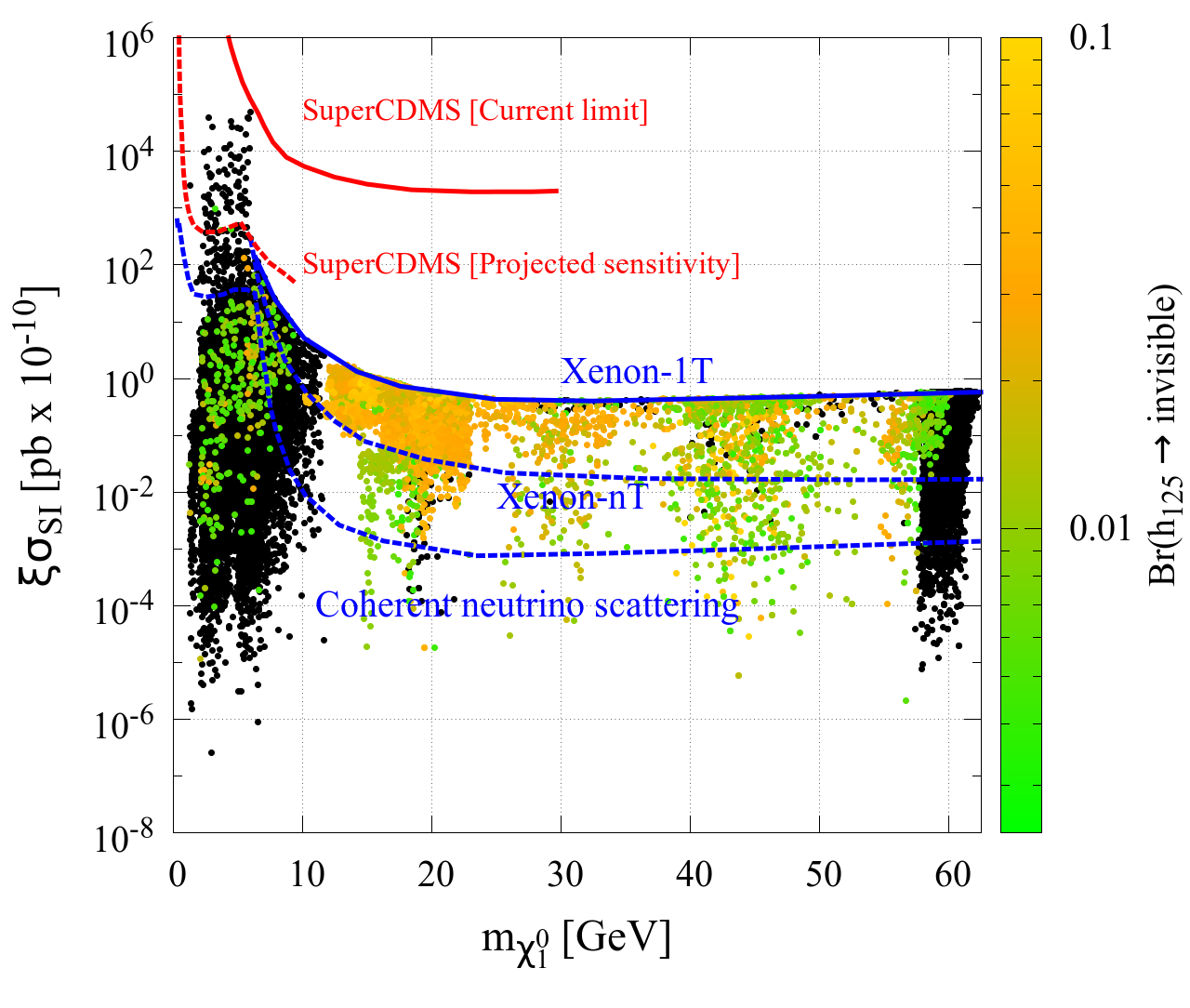}}
\caption{$\xi \sigma_{SI}$ vs $\mlspone$ for the currently allowed parameter space points. \textbf{Left panel}~(modified from Ref.~\cite{Barman:2020vzm})~: The colour palette shows the variation of $\xi$. \textbf{Right panel}~(modified from Ref.~\cite{Barman:2020vzm})~: Black points have $Br(h_{125} \to invisible) < 0.24\%$. The overlapping colour palette shows the variation of $Br(h_{125} \to invisible)$. The reach of various DD experiments is shown by lines labelled accordingly.}
\label{fig:NMSSM_future_DD}
\end{center}
\end{figure}

The allowed parameter space points from \cite{Barman:2020vzm}~(Eqn.~\ref{Eqn:Paramater_scan_NMSSM}) are illustrated in the $\xi \sigma_{SI}$ vs $\mlspone$ plane in the left panel of Fig.~\ref{fig:NMSSM_future_DD} with the colour palette showing the variation of $\xi$. The complementarity between the $\xi \sigma_{SI}$ and the Higgs to invisible branching ratio is also highlighted in the right panel of Fig.~\ref{fig:NMSSM_future_DD}~(taken from Ref.~\cite{Barman:2020vzm}) 
where the black points have $Br(h_{125} \to invisible) < 0.24\%$, and therefore, are outside the projected reach of the CEPC. The overlapping colour palette shows the variation of $Br(h_{125} \to invisible)$ for the allowed points with $Br(h_{125} \to invisible) > 0.24\%$. Here, the Higgs to invisible branching fraction is the sum of multiple decay modes of the $h_{125}$: $h_{125} \to \lspone\lspone, A_1A_1 \to 4\lspone, h_{1} h_{1} \to 4\lspone$ and  $\lsptwo \lspone \to (\lsptwo \to A_1/h_{1} \lspone) \lspone \to 4\lspone$. Note that the CEPC will be able to probe a small fraction of currently allowed points in the $\mlspone \lesssim 10~{\rm GeV}$ region where the projected sensitivity of Xenon-nT falls. The CEPC will also be able to probe the coloured points which fall below Xenon-nT's future reach  in the $\mlspone \geq 10~{\rm GeV}$ region. Furthermore, CEPC will also have reach in the region below the coherent neutrino scattering floor which will be forever outside the projected reach of any future DD experiment. In Fig.~\ref{fig:NMSSM_future_DD}, the solid~(dashed) red line illustrates the current limit~(projected sensitivity) from SuperCDMS~\cite{Agnese:2014aze}~(\cite{Agnese:2016cpb}) at $90\%$ CL. This brings out another complementarity between the DD experiments and the $e^+e^-$ collider experiments. In the small $m_{\lspone}$ region~($\lesssim 7~{\rm GeV}$), a fraction of the parameter space points which  fall outside projected reach of the ILC/CEPC/FCC(ee), have $\xi \sigma_{SI} \gtrsim 5 \times 10^{-44}~{\rm cm^{-2}}$ and are thus within the projected sensitivity of SuperCDMS even if Xenon-nT experiment has no sensitivity there. Reference~\cite{Wang:2020dtb} has also investigated the prospects of Higgs to invisible decay in the allowed region of semi-constrained NMSSM where the gaugino and sfermion mass unification is still maintained. They impose a more stringent constraint on $\Omega_{\lspone} h^{2}$ than in reference~\cite{Barman:2020vzm} and demand further that the $\lspone$ gives the observed relic abundance. This much more constrained analysis too, yields allowed  points, where $\lspone$ pair annihilation is mediated through $h_{1}/A_{1}$ in the region $\mlspone \lesssim 12~{\rm GeV}$. They find that $Br(h_{125} \to invisible)$ is $\lesssim 2\%$ for such points, whereas for allowed points in the $Z$ and $h_{125}$ funnel region the maximum value of $Br(h_{125 } \to invisible)$  is $1\%$ and $0.4\%$, respectively.  

The presence of light $h_{1}$ and $A_{1}$ also offers additional possibilities to probe the currently allowed parameter space via the direct searches for the light Higgs bosons at the future colliders.  
This has been investigated for the HL/HE-LHC in reference~\cite{Barman:2020vzm}.  
After translating  the projection limits from the direct searches in the $h_{125} \to A_{1}A_{1}/h_{1}h_{1} \to 2b2\mu$ channel onto the allowed parameter space, the results indicate that the discovery potential of the HL-LHC as well as the HE-LHC is not very strong. Compared to the projected capability of the HE-LHC, an improvement of more than $2$ orders of magnitude would be required in order to cover the entire allowed parameter space in the $m_{A_1/h_{1}} \gtrsim 15~{\rm GeV}$ region. Note that the projected sensitivity of light Higgs boson searches at the future lepton colliders are projected to be stronger than that of the HL-LHC by around 1-2 orders of magnitude \cite{Liu:2016zki,Drechsel:2018mgd}. In that case, the discovery potential of the future lepton colliders would be even stronger than the HE-LHC.

\begin{figure}[!htb]
\begin{center}
\resizebox{1.00\columnwidth}{!}{
\includegraphics{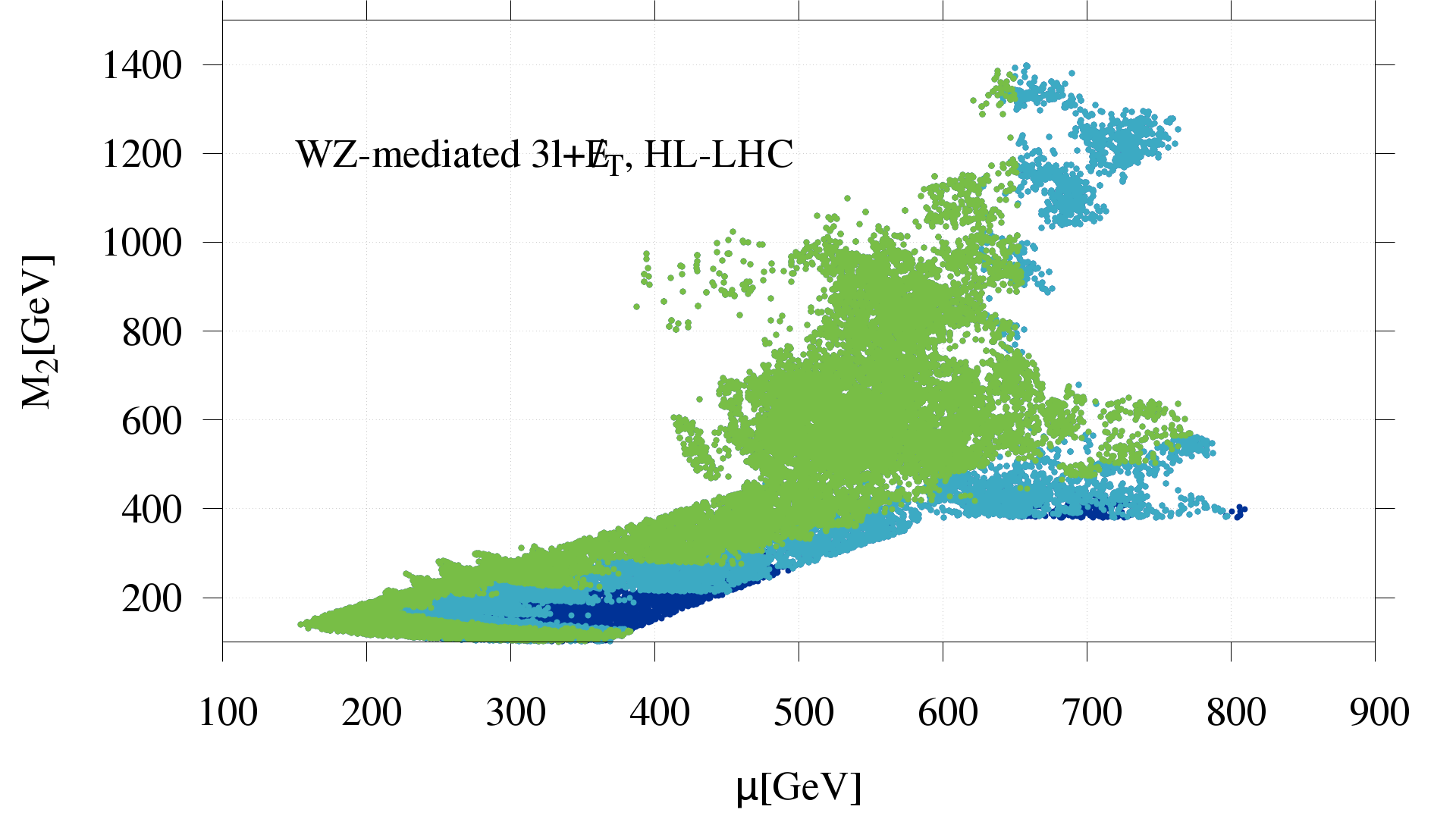}\includegraphics{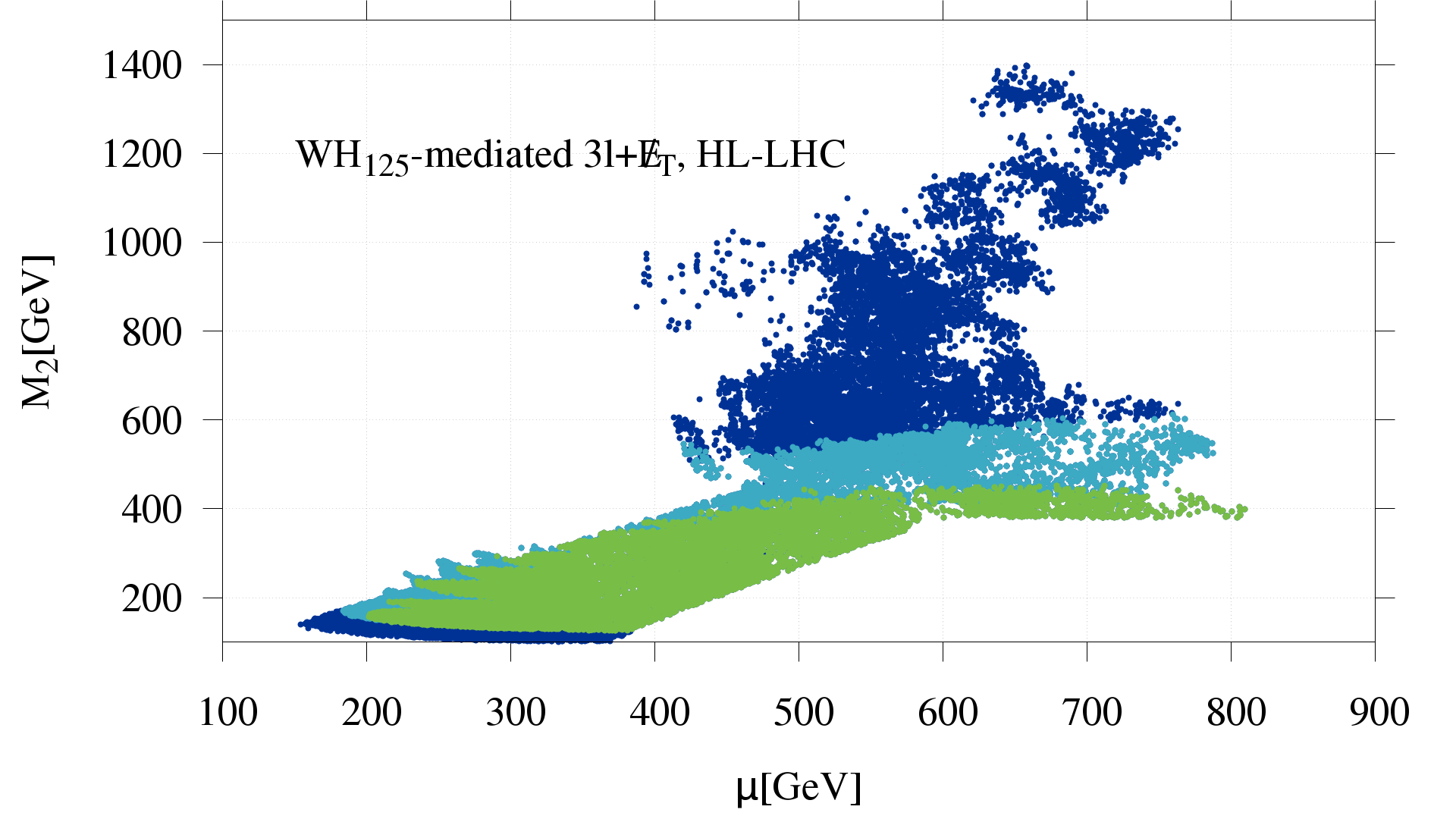}}
\caption{$M_{2}$ vs $\mu$ for the currently allowed parameter space points. The pale blue and green points are within the projected exclusion and discovery reach of direct EWeakino searches in the $WZ$~(left panel) and $Wh_{125}$~(right panel) mediated $3l+\met$ channel at the HL-LHC. The blue points are outside the projected excluded reach of the HL-LHC in the respective search channels~(from Ref.~\cite{Barman:2020vzm}).}
\label{fig:NMSSM_future_HL-LHC}
\end{center}
\end{figure}

The projected reach of direct EWeakino searches in the $WZ$ and $Wh_{125}$ mediated $3l+\met$ channel at the HL-LHC and the HE-LHC has also been analysed in reference~\cite{Barman:2020vzm} and their translation results are illustrated in Fig.~\ref{fig:NMSSM_future_HL-LHC}. The  points shown in Fig.~\ref{fig:NMSSM_future_HL-LHC} correspond to the currently allowed parameter region. The pale blue points and the green points fall within the projected exclusion and discovery reach of direct EWeakino searches in the $WZ$~(left panel) and $Wh_{125}$~(right panel) mediated $3l+\met$ channel at the HL-LHC. The dark blue points have $S_{\sigma} < 2$, thus, falling outside the projected reach of the HL-LHC. At large values of $M_{2}$ and $\mu$, the production cross-section becomes small while the signal efficiency is large. Similarly, in the small $M_{2}/\mu$ region, the production cross-sections are larger but the signal efficiency becomes smaller. This interplay between the production rate and the signal efficiency determines the signal significance of a parameter space point. In Fig.~\ref{fig:NMSSM_future_HL-LHC}~(left panel), dark blue points are observed in the $(\mu~\sim 700~{\rm GeV},~M_{2}\sim 500~{\rm GeV})$ and $(\mu~\sim 300-400~{\rm GeV},~M_{2}\sim 150~{\rm GeV})$ regions. In both these regions, the $M_{2}$ is smaller than $\mu$, and therefore, $\lsptwo$ is dominantly wino in nature. Note that the dominant neutralino-chargino pair production mode is $\lsptwo\chonepm$. Due to small higgsino admixture in $\lsptwo$, the branching fraction of $\lsptwo$ into the $Z\lspone$ final state is suppressed and $\lsptwo$ dominantly decays into $h_{125}\lspone$ if kinematically allowed. As a result, the signal yield of these points in the $WZ$ mediated channel is smaller and thereby, falls outside its projected reach. Both these regions are however within the projected discovery reach of the HL-LHC via direct searches in the $Wh_{125}$ mediated $3l+\met$ channel. The work in reference~\cite{Barman:2020vzm} shows that the direct EWeakino searches in the $3l+\met$ channel at the HL-LHC will be able to probe almost the entirety of the currently allowed parameter space with discovery reach while the same search at the HE-LHC will be able to probe the entire parameter space with much greater signal significance.

\section{Conclusion}

We have reviewed here the status of light~($\leq m_{h_{125}/2}$) LSP in Supersymmetry. Cosmological considerations allow the cold dark matter particle mass to be as low as O(GeV). For the cMSSM, a light thermal DM candidate is all but ruled out. Two possibilities of a light thermal DM in SUSY that are still viable in  variants or extension of the MSSM are: a light neutralino $\lspone$ or a light $\widetilde \nu_R$.

In the pMSSM where the gaugino unification condition is modified or disregarded, there still exist regions of the $M_2$ -- $\mu$ parameter plane, corresponding to $m_{\lspone} \simeq m_{h_{125}}/2~(62.5~{\rm GeV})$, the so called $h_{125}$ funnel region, where the $\lspone$ can be thermal DM and can account for, quite often, at least a substantial fraction, of the observed relic density in the Universe. These  regions in the pMSSM parameter space  are consistent with the current constraints from direct sparticle and BSM Higgs searches at the LHC, as well as from the measurements of mass and couplings of $h_{125}$  and current results from the direct/indirect Dark Matter detection experiments. 
The currently allowed points in the parameter space are clustered around  $\xi \sim 1$ and $\xi \sim 10^{-2}$.
The future DD experiment Xenon-nT will be able to probe the allowed $h_{125}$ funnel region completely. Further confirmations can come from accurate measurements of the invisible width of the Higgs in future $e^+e^-$ experiments.  The HL-LHC~(HE-LHC) will be able to cover partially~(almost completely) the allowed region of parameter space through the searches for EWeakinos through their direct production. 

In the pMSSM analysis if one considers, in addition, the region of parameter space where the relic is overabundant  and assume that  with a non thermal cosmology the $\lspone$ may be  responsible for the observed relic, one sees that the LHC EWeakino  searches and the measurements of the invisible Higgs decays should be able to cover this situation too.  In this case one observes interesting complementarity between the DD experiments, the invisible width measurement of $h_{125}$ at the future $e^+e^-$ colliders and the HL/HE-LHC EWeakino searches. In particular, the current LHC searches for
heavier EWeakinos, rule out regions where the LSP lies in the  mass range
$\lesssim 15$~GeV and which are below the current Xenon-1T sensitivity. In the future,  the HL/HE-LHC EWeakinos searches as well as precise $h_{125}$ invisible width measurements will further probe  the regions at small $\lspone$ masses, which are beyond the reach of current DD experiments.

In the NMSSM, there exist four possible funnel regions, corresponding to the $h_1,A_1,h_{125}$ and $Z$ exchange contributing to the annihilation, where again the $\lspone$ can be a good thermal DM candidate over a wide rangle of $\lspone$ masses all the way down to $\sim 1$~GeV, while satisfying all the current constraints. The lowest values of $m_{\lspone}$  correspond to very light $A_1,h_1$. Future DD experiments (Xenon-nT) and measurements of invisible Higgs decays at a future $e^{+}e^{-}$ collider will be able to cover a significant fraction of the currently allowed parameter space. Here again  the invisible Higgs measurements at the future $e^{+}e^{-}$ colliders will be able to cover the region outside Xenon-nT's future reach over the entire $\mlspone$ range. Moreover,  the HL-LHC and HE-LHC will be able to cover almost the entire region of the currently allowed parameter space via the direct searches of EWeakinos in the $3l+\met$ channel with discovery reach.

Finally light sneutrinos provide an alternate DM candidate. A $\widetilde \nu_L$ with mass greater than 55 GeV is viable only if another DM particle is responsible for the observed relic. However, it is consistent with the current LHC DM searches only for compressed spectra. In extensions of the SUSY models which also contain a $\widetilde \nu_R$,  the case of a pure $\widetilde\nu_R$, with mass $\sim 30$ -- $40$ GeV, as thermal DM, has been shown to be viable in presence of a quasi-stable $\widetilde \tau$ and is consistent with LHC constraints. In extensions of NMSSM, it is possible to realise a pure $\widetilde \nu_R$  LSP, which is light, can be thermal relic and is consistent with the current constraints.  All these scenarios will give rise to very distinctive phenomenology and can be tested at the HL-LHC.

\vspace{1cm}

{\bf \noindent Acknowledgements :}

The work of RMG is supported by the Department of Science and Technology, India under Grant No. SR/S2/JCB-64/2007.

\vspace{1cm}

{\bf \noindent Author contribution:}

All authors have contributed equally.  

%

\end{document}